\documentclass[secnumarabic,amssymb,superscriptaddress, aps, pre]{revtex4}

\usepackage{amsmath}
\usepackage{mathtools}
\usepackage{graphicx}
\usepackage{xcolor}
\usepackage{epsfig}
\usepackage{mathrsfs}
\usepackage{amsbsy}
\usepackage{verbatim}
\usepackage[colorlinks,citecolor=blue]{hyperref}%

\usepackage[FIGTOPCAP]{subfigure}
\usepackage[subfigure]{tocloft}
\usepackage{latexsym,oldgerm}
\usepackage{longtable}
\def\be{\begin{equation}}
\def\ee{\end{equation}}
\def\bea{\begin{eqnarray}}
\def\eea{\end{eqnarray}}
\def\nn{\nonumber}

\begin{document}
\title{Cell nucleus as a microrheological probe to study the rheology of the cytoskeleton }
\author{Moslem Moradi}
\email[Email: ]{mmoradi@email.unc.edu}
\author{Ehssan Nazockdast}
\email[Email: ]{ehssan@email.unc.edu}
\affiliation{ Department of Applied Physical Sciences, University of North Carolina at Chappel Hill, Chapel Hill, NC 27599-3250}

\date{\today}
\begin{abstract}
\vspace*{0.55cm}
\subsection*{Abstract}
\noindent 
Mechanical properties of the cell are important biomarkers for probing its architectural changes caused by cellular processes and/or pathologies. 
The development of microfluidic technologies have enabled measuring cell mechanics at high-throughput, 
so that mechanical phenotyping can be applied to large samples in reasonable time-scales. 
These studies typically measure the stiffness of the cell as the only mechanical biomarker, 
and cannot disentangle the rheological contribution of different structural components of the cell, 
including the cell cortex, the interior cytoplasm and its immersed cytoskeletal structures, and the nucleus. Recent advancements in high-speed fluorescent imaging have enabled probing the deformations of the cell cortex, while also tracking different intracellular 
components in rates applicable to microfluidic platforms. 
We present a novel method to decouple the mechanics of the cell cortex and the cytoplasm 
by analyzing the correlation between the cortical deformations that are induced by external microfluidic flows, and the 
nucleus displacements induced by those cortical deformations; \textit{i.e.} we use the nucleus as a high-throughput microrheological probe 
to study the rheology of the cytoplasm, independent of the cell cortex mechanics. 
To demonstrate the applicability of this method, we consider a proof of concept model consisting of a rigid spherical nucleus centered in a spherical cell. 
We obtain analytical expressions for time-dependent nucleus velocity as a function of the cell deformations, 
when the interior cytoplasm is modeled as a viscous, viscoelastic, porous and poroelastic materials, and demonstrate how the nucleus 
velocity can be used to characterize the linear rheology of the cytoplasm over a wide range of forces and time-scales/frequencies.
\end{abstract}
\maketitle
%\tableofcontents

% keywords can be removed

\section{Introduction}
Extensive research over the past few decades have demonstrated the key role of the cell's mechanical aspects 
in many cellular processes \cite{fletcher2010cell}, including mechanotransduction \cite{jaalouk2009mechanotransduction}, cell motility \cite{jaalouk2009mechanotransduction}, division \cite{dumont2009force, redemann2020integrated} and differentiation \cite{altman2002cell}. 
The ability of the cell to generate and respond to forces is controlled by its cytoskeleton: 
a highly dynamic and heterogeneous assembly of biopolymers, crosslinkers and motor proteins \cite{howard2001mechanics}.  
Moreover, many malignancies involve large variations in the the cytoskeletal and nuclear organizations,  
which naturally lead to variations in cell mechanics. Thus, cell mechanics provides a label-free 
biomarker for determining cell types, the cell cycle stage, division rate, and disease state among other things \cite{diCarlo2012mechanical, Darling2015}. 

The main structural components that determine the mechanics of eukaryotic cells are: ($\mathrm{i}$) 
The cell membrane and the cortical network of actin filaments attached to it (referred to as the cell cortex hereafter); 
($\mathrm{ii}$) the cell nucleus composed of the nuclear membrane and the nucleoplasm within; 
($\mathrm{iii}$) the cell cytoplasm that fills the volume between the cell cortex and 
the cytoskeletal assemblies within the cytoplasm. These 
include the microtubule asters bound to the nuclear membrane, 
and the immersed intermediate filament networks. 
In this work, we refer to the cytoplasm and its embedded structures as simply cytoplasm 
and refer to the cytoplasmic fluid as cytosol.

The cell's mechanical response to external forces and boundaries 
is determined by force transport within each of these structures and across their interfaces.
%and is a key component in mechanotransduction and mechanosensing processes. 
Indeed, a large number of careful experiments suggest that the cell cortex and the
cytoplasm have two distinct time-dependent mechanical behaviors \cite{hoffman2006consensus,jalai2016parallel}, and 
that the cell nucleus is the stiffest component between these three structures \cite{Dahl2004}. 
We also know that different malignancies and cellular processes may have different effects on these three components. 
For example, taxol is a drug used in cancer therapy that acts through the disruption of 
microtubule assembly and cell division within the cytoplasm \cite{yvon1999taxol}. On the other hand,  
the softness of cancerous cells is associated with the disruption of the cortical actin network \cite{swaminathan2011mechanical}, 
while many diseases involve changes in nuclear architecture and shape \cite{Darling2015}. 
Thus, to explore these structural changes, it is essential to have a measure of the 
relative contribution of each of these structures to the overall mechanical response 
in the experimental techniques.  

The probe-based (local) techniques --including Atomic Force Microscopy (AFM)\cite{krieg2019atomic}, 
passive particle tracking microrheology, and active microrheology using optical tweezers, 
and magnetic particles \cite{Wirtz2009}-- have the advantage of exploring the local (cell cortex, cytoplasm, and the nucleus) mechanical 
response over a wide range of forces and frequencies (time-scales). 
These methods are, however, low-throughout (less than $100$ cell per hour).
Cells of different type, size, cell cycle stage, and malignancies can have 
significantly different cytoskeletal and nuclear organizations and mechanical properties.  
This makes applying these methods to large and heterogeneous population of cells very challenging. 

In contrast, serial deformations of cells in microfluidic platform through hydrodynamic forces enables 
throughput rates of $100-1000$ cells per second, which makes them suitable for clinical applications.  
Gossett \textit{et al.}, combined high speed imaging with cell deformation in extensional flows and used the ratio of the longest (stretched) to shortest (compressed) axes of the cells as a measure of cell of deformability in extension flows to separate diseased and healthy cells \cite{gossett2012hydrodynamic}. Mietke \textit{et al.} used the analytical solution 
of the flow around a spherical cell in an axisymmetric cylindrical channel to compute 
the fluid stresses on the cell and compute steady-state deformation of cell, assuming small deformations and modeling the cell as an elastic shell or sphere \cite{mietke2015extracting}. 
The predicted shapes were compared against the cell shapes from microscopy  to extract the elastic properties of the cell 
in both bulk and shell models. 
Mokbel \textit{et al.} used numerical simulations to extend these calculations to finite deformations, 
where they modeled the cell as a viscoelastic material surrounded by a thin elastic shell \cite{mokbel2017numerical}. 
Most recently, Fregin \textit{et al.} used dynamic real-time deformability cytometry to simultaneously measure the \emph{apparent} elastic and viscous properties of cell without solving for the detailed fluid flows and deformation fields \cite{fregin2019high}. 

One limitation of these studies is that they do not explicitly account for the nucleus in their mechanical models. 
However, we know that the nucleus to the whole cell size ratio --which is a purely geometrical factor-- can largely change 
the apparent stiffness and fluidity of the cell  \cite{ribeiro2010nucleus, darling2011force}.

Another limitations of these studies is that they have not explored the time-dependent mechanics of 
the cytoplasm \cite{mietke2015extracting, fregin2019high}. 
However, previous microrheological measurements show 
that the elastic and viscous responses of the cytoplasm and its 
immersed cytoskeleton is generally time-dependent \cite{mofrad2006cytoskeletal,fabry2001scaling}. 
Because of the highly dynamic and hierarchical structures immersed in cytoplasm,  its rheology  
can vary significantly with time- and length-scale of the measurements \cite{pullarkat2007rheological,janmey2007cell,
trepat2008universality,mahaffy2004quantitative,
cai2013quantifying,cartagena2014local}.
 This complexity in structure and mechanical response has also led to several competing theories for describing the rheology of the cytoplasm. In continuum limit, the cytoplasm has been modeled as a viscoelastic material, 
a porous gel,  an active gel and a glassy material  \cite{lim2006mechanical,yeung1989cortical,strychalski2016intracellular,
alt1999cytoplasm,gracheva2004continuum,moeendarbary2013cytoplasm}. 

Recent technologies in fluorescent and high speed imaging have enabled simultaneous tracking of the time-dependent shape 
of the cell cortex, and displacements of different intracellular components particularly the nucleus, 
with high spatial and temporal resolution as the cells flow through the microchannels \cite{diebold2013digitally, mikami2018ultrafast, rosendahl2018real}. 
These advancements offer the opportunity to build upon the previous theories and experiments and devise techniques for
high-throughput time-dependent mechanical characterization of different cellular components.  
Yet, the modeling framework for deploying this rich data is underdeveloped.   

The aim of this study is to use the time-dependent deformation of the cell cortex and the displacement of the cell nucleus 
to determine the time-dependent mechanics of the cytoplasm independent of the cortex mechanics. 
Specifically, we use the nucleus as a microrheological probe to study the cytoplasm mechanics. 
Unlike the typical microrheology experiments where the probe's motion is driven by thermal forces 
(passive microrheology) or external forces (active microrheology), in our system the motion of the nucleus 
is driven by the deformations of the cell boundary (cortex), which itself is induced by external microfluidic flows. 

Given that the nucleus is significantly stiffer than the cell cortex, in the first step, we model the nucleus as a rigid sphere. 
Assuming the cortex deformations are known through experiment, the nucleus velocity 
is entirely determined by the rheology of the cytoplasm and the size of the nucleus, irrespective of the mechanics of the cortex. 
To make analytical progress we consider a proof of concept model consisting of a 
spherical nucleus centered in a spherical cell and provide 
analytical solutions to the cytoplasm velocity field and the nucleus velocity 
as a function of the rate of displacement of the cortex at small deformations. 
We model the cytoplasm as viscous, porous, viscoelastic, and poroelastic media and explore the effect of the constitutive parameters, such as relaxation time in Maxwell equation and permeability and network relaxation time in poroelastic models,
in the time-dependent relationships between cell cortex surface velocity modes and the ensued nucleus velocity.

\section{Formulation and assumptions}
\begin{figure}[!htb]
    \centering
    \subfigure[]{\includegraphics[width=0.62\textwidth , height=4.7cm ]{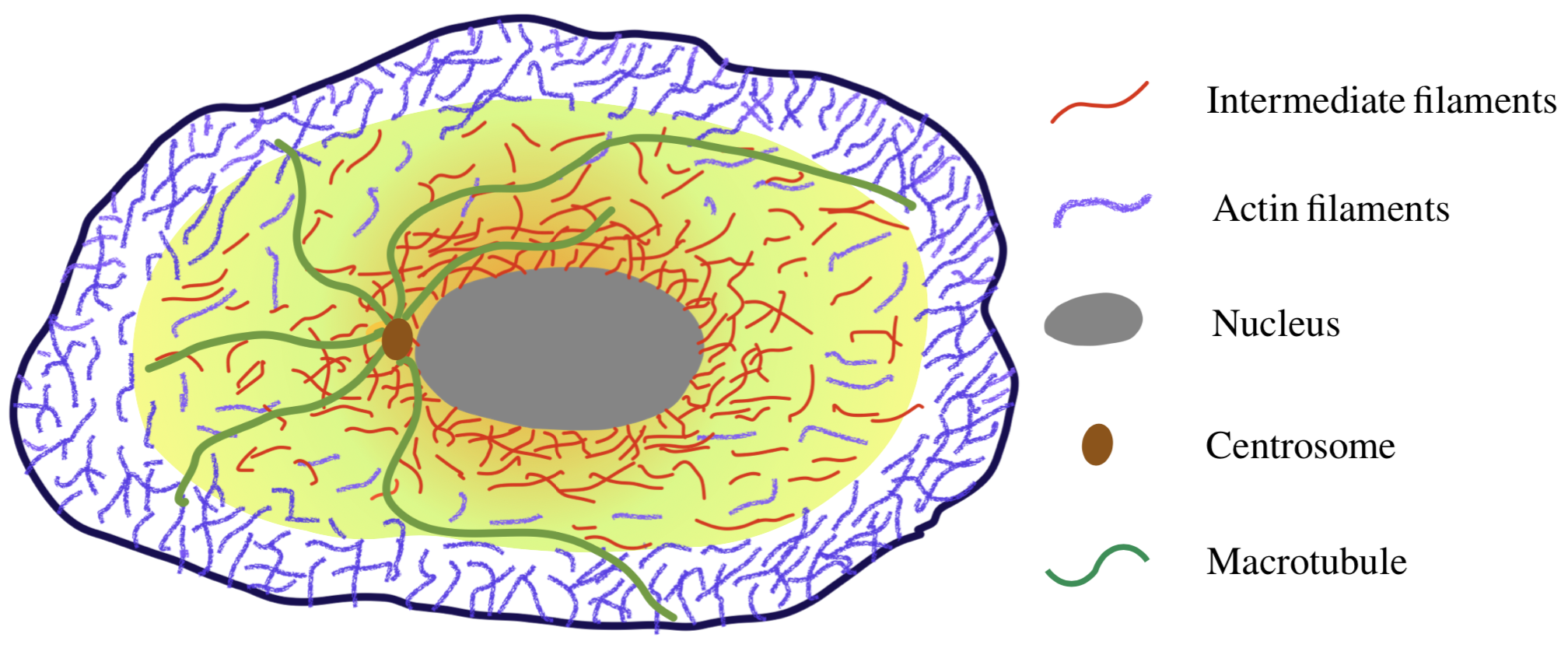}} \label{Fig1a}
    \quad
    \subfigure[]{\includegraphics[width=0.35\textwidth , height=4.7cm]{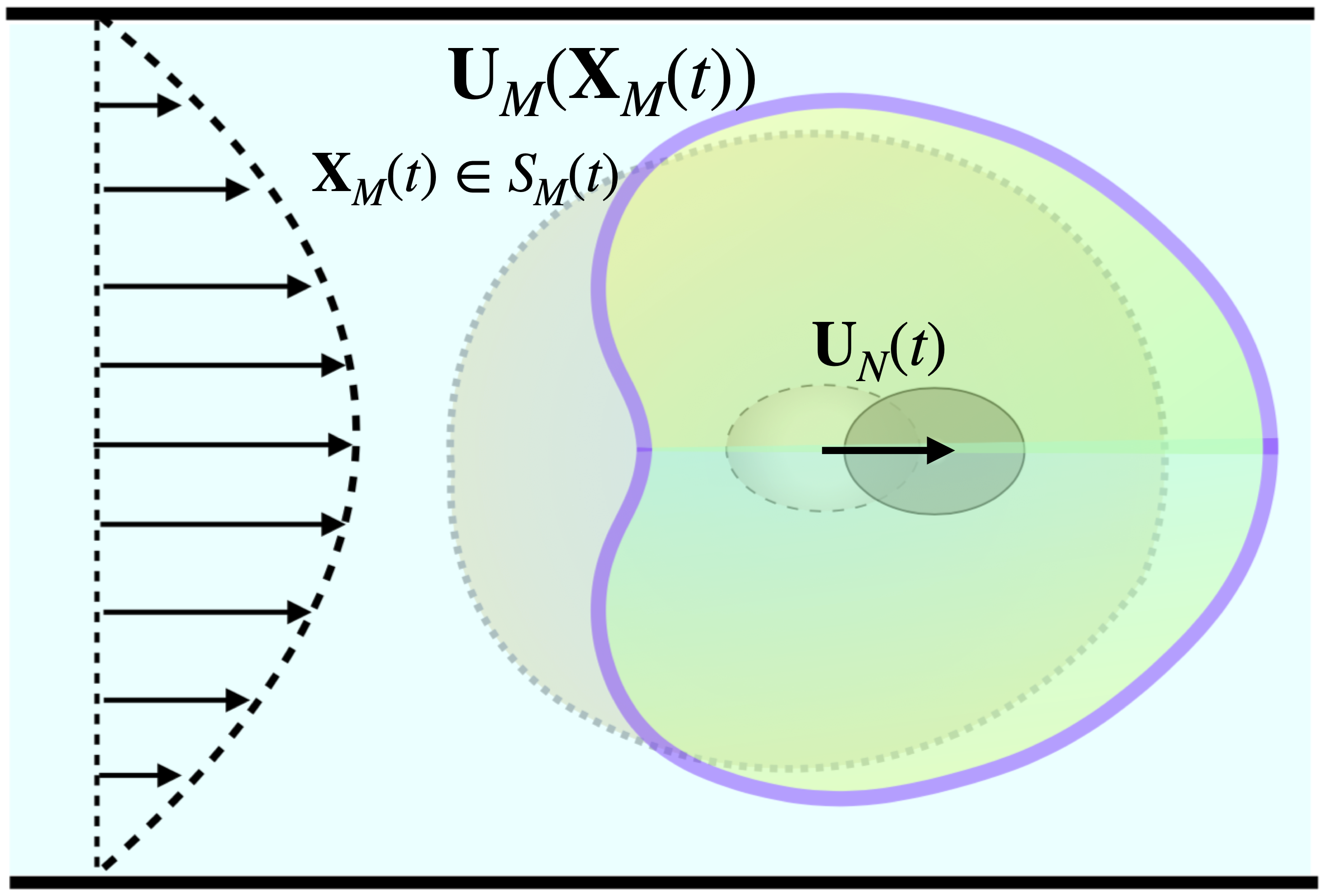}} \label{Fig1b}
    \caption{(a) A schematic representation of the intracellular assemblies formed by 
    cytoskeletal filaments, namely microtubules, actin filaments 
    and intermediate filaments. The flexible filaments are either \emph{freely suspended} or \emph{crosslinked} 
    and anchored to the nucleus. When the filaments are freely suspended the cytoplasm behaves as a viscoelastic fluid in the continuum scale. When the filaments are crosslinked, they form a network and behave as a poroelastic material in the continuum scale. (b) A schematic representation of the time-dependent deformations of the cell cortex under external hydrodynamic forces from pressure-driven flows in mirochannels and the subsequent displacements of the nucleus, induced by the internal flows generated through the deformations of the cell cortex. Assuming the cortical deformations and hence the surface velocity, $\mathbf{U}_\text{M}$, are known through experiments,the nucleus velocity, $\mathbf{U}_N$, can be decoupled from cortex mechanics, and fully determined by the surface velocity for given constitutive equation for the cytoplasm.} 
    %The force exerted on the membrane by the outer fluid,$\mathbf{f}_{\text{O}}$, and forces induced by interior cytoplasm, $\mathbf{f}_{\text{I}}$, is balanced by forces induced by membrane deformation, $\mathbf{f}_\text{M}$.}
    \label{fig1}
\end{figure}

Cell deformations under external flows are determined by the balance between the forces exerted on 
the cortex from the outer fluid, ${\mathbf{f}}_{\mathrm{O}}$, forces induced by the interior cytoplasm, 
${\mathbf{f}}_{\mathrm{I}}$, and the forces induced by the cortex deformations, 
${\mathbf{f}}_{\mathrm{M}}$. In the absence of inertial forces in Stokes regime, the sum of these forces is identically zero: 
${\mathbf{f}}_{\mathrm{O}}+{\mathbf{f}}_{\mathrm{I}}+{\mathbf{f}}_{\mathrm{M}} =0$.
Assuming we can track the cell surface, one can compute the outer fluid flow and its  
associated forces by solving the Stokes flow and imposing a no-slip boundary condition  on the cell's outer surface; see Fig.\ref{fig1}(b).
In the presence of a constitutive model for the cortex, ${\mathbf{f}}_{\mathrm{M}}$ can be computed as a function 
of its time-dependent deformation. Hence, we can compute 
${\mathbf{f}}_{\mathrm{I}}= -( {\mathbf{f}}_{\mathrm{O}}+{\mathbf{f}}_{\mathrm{M}} )$, 
which can be used to extract the coefficients of the presumed form of constitutive equation for the cell cytoplasm. 
In other words, determining the rheology of the interior cytoskeleton requires specifying the constitutive equation for the cortex.

Here we propose an alternative route that allows determining the rheology of the cytoplasm independent
of the cortex mechanics. The main assumption here is that the cell's surface velocity is known through experiments. 
We use this surface velocity as a boundary condition (BC), while BCs for nucleus is no-slip and zero net force.  
This provides sufficient conditions for computing the internal cytoplasmic flows, when complemented  with 
the appropriate form of momentum equation 
for viscous, viscoelastic, porous and poroelastic materials. 
Note that these calculations are completely decoupled from the cortex mechanics. 

The remaining assumptions, enumerated below, are made for simplicity and to allow analytical progress. 
The methodology presented here can still be applied even when these assumption are relaxed, but the
equations of motion may need to be solved numerically.  These assumptions are (i) the nucleus is modeled as a rigid sphere , and (ii) is centered in 
a spherical cell; (iii) the deformations of the cell boundary and the net displacement of the nucleus are assumed to be small, so that 
the geometry remains concentric. In these conditions, the flow within the cytoplasm is axisymmetric in spherical coordinates.   
\section{Results }
In the next several sections, we present our analytical results of the nucleus velocity as a function of the rates of deformations of the cell cortex, when the cytoplasm is a Newtonian fluid, a viscoelastic fluid, a porous material and a poroelastic network. 
To facilitate the use of these often very long 
analytical expressions and for further clarity, 
we reproduce the detailed steps taken to obtain the final expressions in each section in an accompanying \textit{MATLAB Live Script} --where we use  
MATLAB symbol toolbox for manipulating the analytical equations. 
\subsection{Nucleus surrounded by a Newtonian fluid}
\begin{figure}
    \centering
    \subfigure[]{\includegraphics[width=0.28\textwidth , height=5cm ]{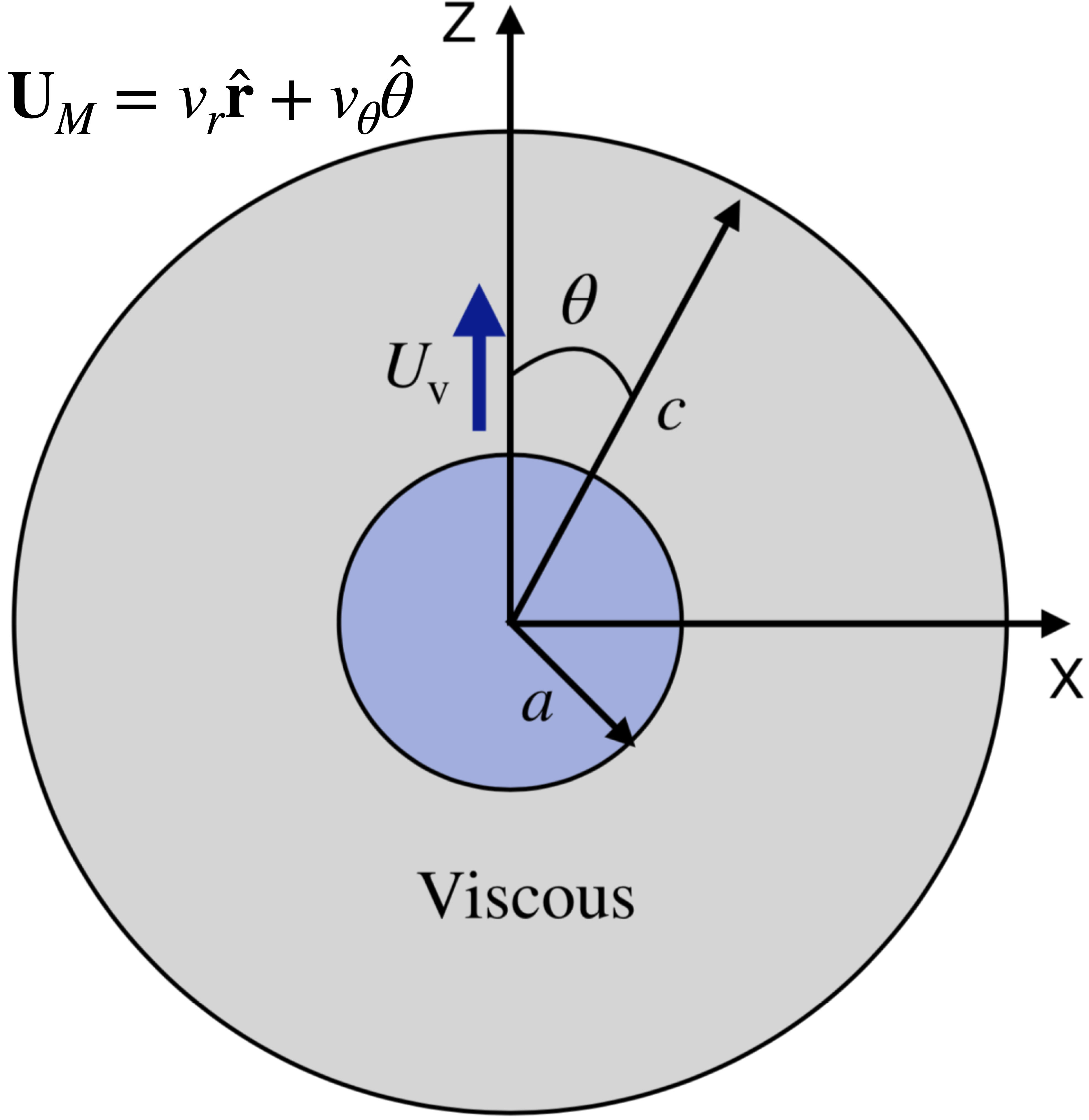}} \label{Fig2a}
    \subfigure[]{\includegraphics[width=0.28\textwidth , height=5cm]{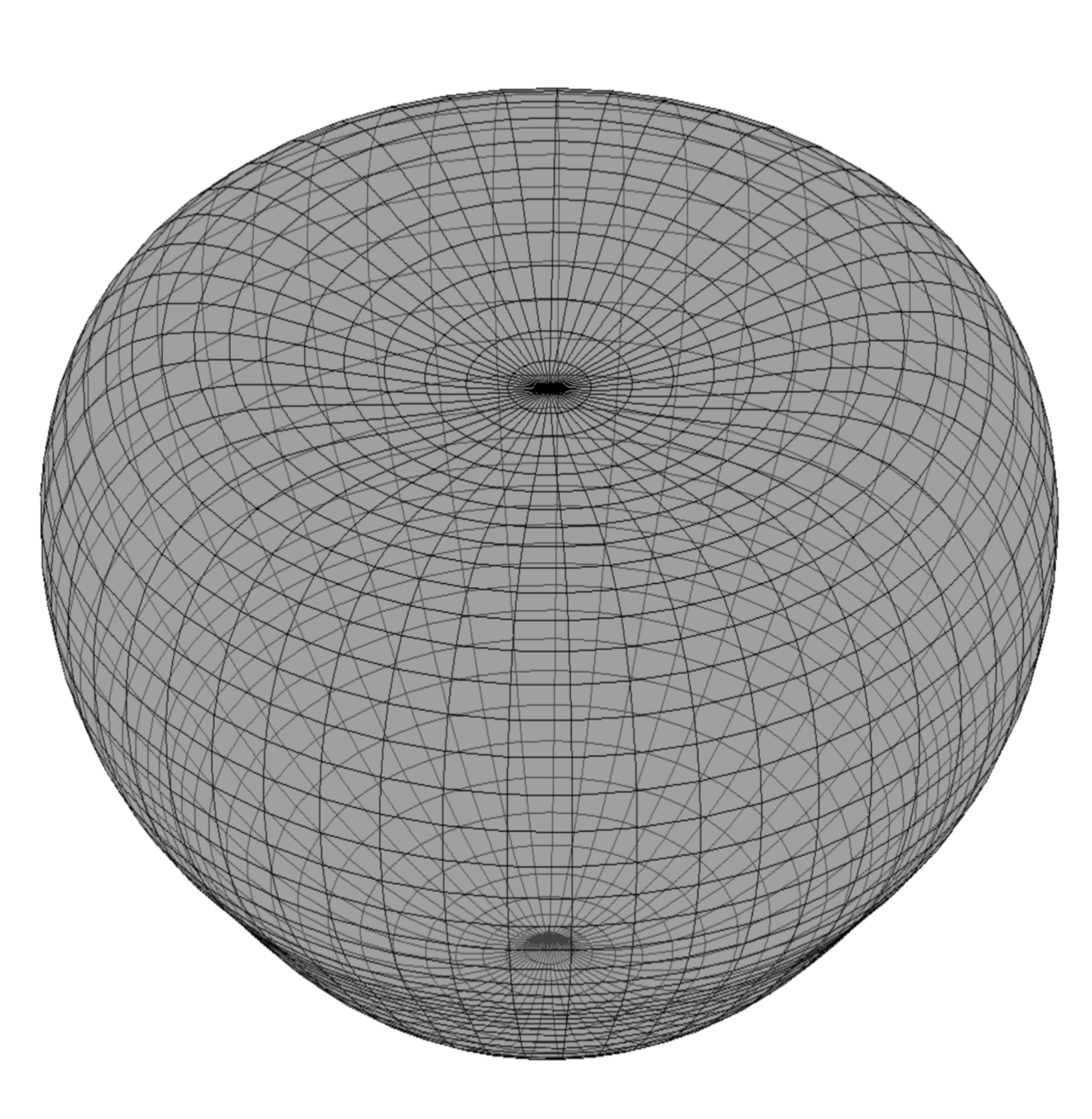}} \label{Fig2b} \quad
    \subfigure[]{\includegraphics[width=0.4\textwidth , height=5.5cm]{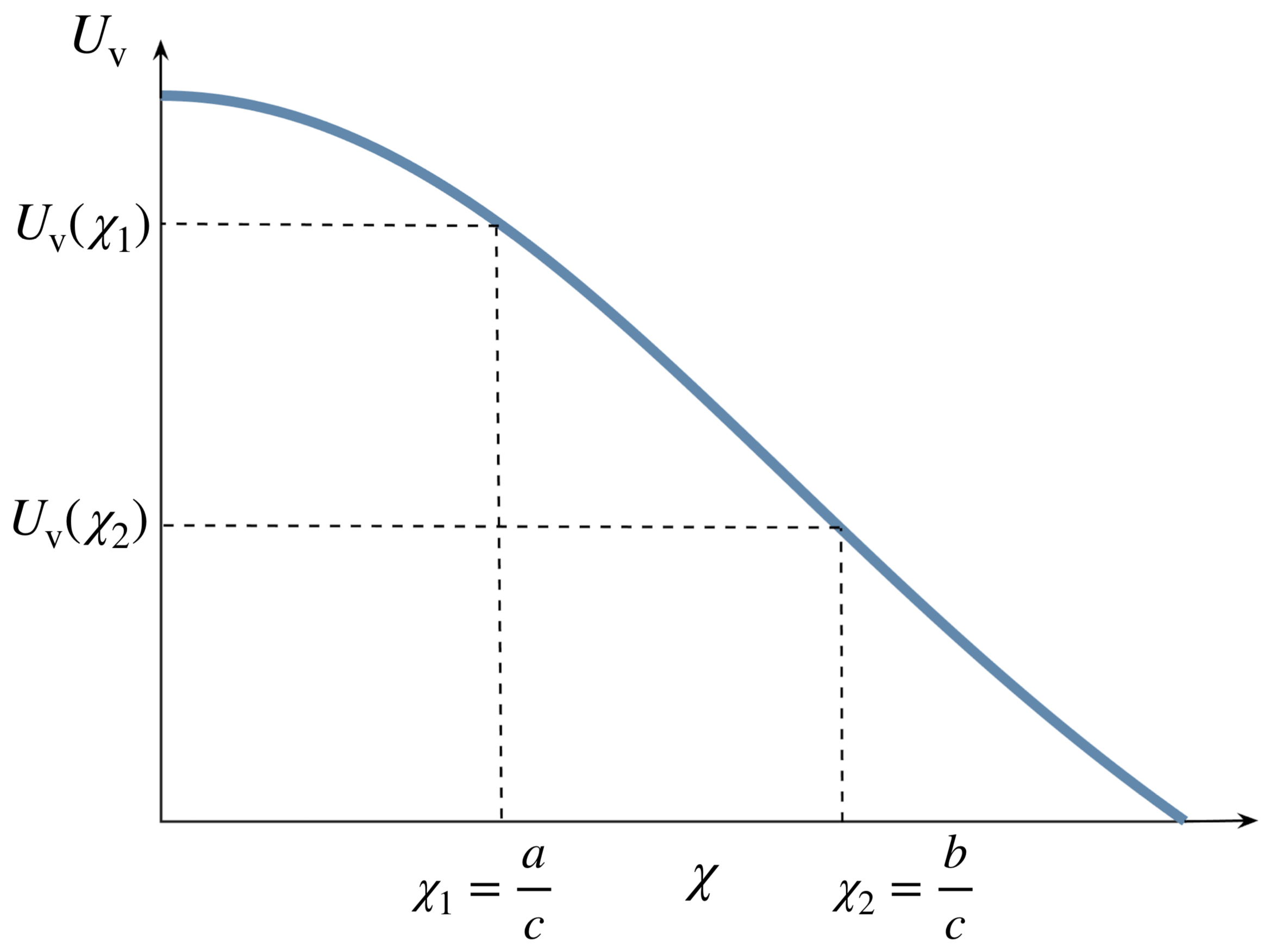}} \label{Fig2c}
    \caption{(a) A force-free spherical nucleus moving with velocity $U_{\mathrm{v}}$ in Newotonian fluid  and confined within a spherical cortex of radius $c$. Due to axisymmetry, the cortex rate of deformations can be decomposed into radial and angular modes of rates of displacement: $v_r=\sum_{n=1}^\infty {\lambda}_n P_n(\cos \theta)$ and $v_\theta=\sum_{n=1}^\infty {\lambda}^\prime_n V_n(\cos \theta)$, where $V_{n} (\cos\theta) = \frac{2 \sin\theta}{n(n+1)} \frac{\mathrm{d} P_n (\cos\theta)}{\mathrm{d} \cos\theta}$.
    (b) Outer cell shape for some special values of the first five modes that recapitulates a bullet shape cell: ${\lambda}_n =[1.0000  ; -1.1975 ;  -0.8497  ; -0.8296  ; -0.2806]$ and $ {\lambda}_{n}^{\prime} = [1.0000  ;  1.7277  ;  1.3449 ;  -0.8401 ;  1.8323 ]$. 
    (c) Velocity of the inner sphere as a function of the nucleus to the cell cortex radii, $\chi =a/c$, when surface velocity modes are ${\lambda}_1 = {\lambda}_{1}^{\prime} =1$.
     }
    \label{fig2}
\end{figure}

We start by the simplest mechanical model and assume the cytoplasm is a Newtonian fluid of shear viscosity $\eta_\circ$.
The nucleus is modeled as a rigid sphere of radius $a$ centered  in a spherical cell of radius $c$; see Fig.\ref{fig2}(a).
 Since the flow is axisymmetric, the velocity field in spherical coordinates are $\mathbf{v}(r,\theta )= (v_r , v_{\theta})$, with $\theta\in[0,\pi]$. 
 The BCs are no-slip on cell outer surface, $\mathbf{v}(r=c,\theta )$, and the net velocity of the cell nucleus is $U_{\mathrm{v}}$ on the inner surface, $\mathbf{v}(r=a,\theta )$, where "$\mathrm{v}$" in $U_{\mathrm{v}}$ stands for viscous.  Both of these quantities are inputs from experiments. 
 The solutions to radial and angular velocity fields in axisymmetric Stokes flow are given in terms of 
Legendre polynomials and their derivatives; see Appendix~\ref{appA}. 
Thus, we present the outer surface velocity using the same functions.  
 The additional equation needed to complete the system of equations is the constraint that the nucleus is force-free.  
 These conditions are given by the following equations: 
 \begin{subequations}
\begin{align}
r&=a: & &v_r= U_{\mathrm{v}} \cos\theta,  && v_{\theta}= - U_{\mathrm{v}} \sin\theta, & &\int_{S_a} \left(\mathbf{\sigma}\cdot \mathbf{n}\right)\cdot \mathbf{\hat{z}} dS=0.&\\ 
%& v_{r}{\Big|}_{r=a} = U \cos\theta , & \quad \quad & v_{\theta}{\Big|}_{r=a} = - U_1 \sin\theta,  \\ 
 r&=c:  & &v_{r}  = \sum_{n=1}^{\infty} {\lambda}_n P_n (\cos\theta), & &v_{\theta} = \sum_{n=1}^{\infty} {\lambda}_{n}^{\prime} V_n (\cos\theta),&
%& \int_{S_a} \left(\mathbf{\sigma}\cdot \mathbf{n}\right)\cdot \mathbf{\hat{z}} dS=0
\end{align}
\label{eq:BC1}
\end{subequations}
where $\displaystyle{ V_{n} (\cos\theta) = \frac{2 \sin\theta}{n(n+1)} \frac{\mathrm{d} P_n (\cos\theta)}{\mathrm{d} \cos\theta} }$, and $P_n (\cos\theta)$ is the Legendre polynomial of order $n$. Note that the choice of $P_n(\cos \theta)$ and $V_n(\cos \theta)$ for radial and angular velocities readily 
satisfies the incompressibility constraint, $\nabla \cdot \mathbf{v}=0$. 

%We use general stream function solution of Stokes equation for axisymmetric geometries (see Appendix~\ref{appA}).  
After applying the BCs the velocity of the inner sphere is given by:
\begin{eqnarray}
&& U_{\mathrm{v}} =  \frac{1}{3} {\lambda}_1 \left(  \frac{6+ 6\chi + 
{\chi}^2 +  {\chi}^3 +  {\chi}^4}{1+ \chi + {\chi}^2 + {\chi}^3 + {\chi}^4} \right) - 
\frac{1}{3} {\lambda}_{1}^{\prime} \left(  \frac{-3- 3 \chi +2 {\chi}^2 +2 {\chi}^3 +2 {\chi}^4}{1+ \chi + {\chi}^2 + {\chi}^3 + {\chi}^4} \right),  
\label{U1solid} 
\end{eqnarray}
where $\chi = \frac{a}{c}$. This velocity depends on just the first modes of cortex deformation rates 
and independent of the cytoplasmic viscosity. 
Figure \ref{fig2}(c) shows the velocity of the inner sphere as function of $\chi$. 
As a numerical example, if we take ${\lambda}_1 = {\lambda}_{1}^{\prime} =1$, for $\chi = \frac{1}{3}$, 
the velocity of the inner sphere is $\frac{959}{363} \approx 2.64$, 
for $\chi =\frac{2}{3}$ it is $U_{\mathrm{v}} = \frac{1139}{633} \approx 1.79$, and for $\chi\rightarrow 0$ it is $U_{\mathrm{v}} = 2 {\lambda}_1 + {\lambda}_1^{\prime}=3$.

As we shall see in the next sections, because of the flow axisymmetry the nucleus velocity  
only depends on the first surface velocity modes of the cortex, $\lambda_1$ and $\lambda^\prime_1$,
regardless of our choice of model for cytoplasm.

\subsection{Viscoelastic cytoplasm}
Particle-tracking microrheology has been extensively used to measure the linear viscoelastic response of the cytoplasm for 
many varieties of cells under a wide range of conditions; see \cite{Wirtz2009} and references within for a review on the topic. 
These studies show that in many cases the cytoplasm exhibits frequency-dependent 
storage and loss moduli \cite{Wirtz2009,mofrad2006cytoskeletal,fabry2001scaling}. 
One commonly observed behavior is the power-law dependence of the storage and loss moduli. 
Soft glassy rheology model is used to describe this behavior \cite{bouchaud1992weak,sollich1997rheology}.
However, the type of constitutive equation for a given cell type is generally not known. 
Thus, we use the most general description of a linear viscoelastic incompressible fluid given by 
\be
\boldsymbol{\sigma}^{p}(t)=G(t)\boldsymbol{\gamma}(0)+ \int_0^t G(t-t^\prime)\boldsymbol{D} (t^\prime)\, \mathrm{d} t^\prime
\ee
where superscript $p$ stands for polymer (filament) phase, $\boldsymbol{\sigma}(t)$ is the stress tensor, 
$\boldsymbol{\gamma}(t)$ is the strain tensor, 
$\mathbf{D} (t) =\partial \boldsymbol{\gamma}/\partial t= \nabla \mathbf{v} + \nabla {\mathbf{v}}^{\mathrm{T}}$ is the time-dependent rate of strain tensor, 
and $G(t)$ is the time-dependent shear modulus. 
The goal is to determine $G(t)$ as a function of the cell surface and the nucleus velocities.  
 
Intracellular structures in eukaryotic cells are highly heterogeneous. For example, the density of microtubules and intermediate filaments 
are typically largest near the nucleus and decays towards the outer boundaries; see Fig.\ref{fig1}(a).  
We model these structural heterogeneities by decomposing the cytoplasm into two domains: we assume that the nucleus of radius $a$ is bound by a spherical shell of outer radius $b$ that contains the viscoelastic fluid. This shell is bounded by a Newtonian fluid that fills the remaining intracellular volume; see Fig.\ref{fig4}(a). Given that cytoskeletal filaments can be tagged and visualized in microchannels  \cite{diebold2013digitally, mikami2018ultrafast, rosendahl2018real}, one can analyze the intensity of the fluorescent to estimate the density distribution of the filaments and approximate $b$.  

The total stress within the shell, $a <r<b$ is the sum of the polymeric and cytosol stresses: 
$\boldsymbol{\sigma}=\eta_\circ \mathbf{D}+\boldsymbol{\sigma}^{p}$. Taking Laplace transform of the total stress in time axis and using convolution theorem gives: 
\be
\tilde{\boldsymbol{\sigma}}(s,r,\theta)=\left(\eta_\circ+\tilde{G}(s)\right) \left(\nabla \tilde{\mathbf{v}}(s,r,\theta)+\nabla \tilde{\mathbf{v}}^T(s,r,\theta)\right),
\label{eq:stressLaplace}
\ee
 where superscript $\sim$ denotes variables in Laplace space. Taking Laplace transform of the momentum equation in the shell and substituting 
 for stress with Eq.\eqref{eq:stressLaplace} results in Stokes equation in Laplace space
 \be 
 \tilde{\eta} {\nabla}^2 \tilde{\mathbf{v}} - \nabla \tilde{p} = 0, \qquad\qquad a<r<b \label{eq:StokesLaplace}
\ee
where $\tilde{\eta} =\left( \eta_{\circ}+\tilde{G}(s) \right)$. 
Similarly, the momentum equation for outer Newtonian fluid is another Stokes equation with $\eta_{\circ}$ as the shear viscosity. 
The BCs at the interface, $r=b$, are the continuity of fluid velocity, tangential stress and pressure, given bellow:
\begin{align}
&r=b: & &\tilde{\mathbf{v}}^{\mathrm{(I)}} = \tilde{\mathbf{v}}^{\mathrm{(II)}},
&& \tilde{\sigma}_{r\theta}^{\mathrm{(I)}} =\tilde{\sigma}_{r\theta}^{\mathrm{(II)}}, && \tilde{p}^{\mathrm{(I)}}= \tilde{p}^{\mathrm{(II)}}.       
\label{eq:BC2}
\end{align} 
  
We, then, use the stream function solution to Stokes flow in domains $\mathrm{I}$ and $\mathrm{II}$ 
and apply BCs in  Eqs.\eqref{eq:BC1} and \eqref{eq:BC2} to 
analytically determine $\tilde{U}(s)$ as a function of surface velocity modes at $r=c$ as
\be\tilde{U}(s)= \frac{1}{3\Delta} \left[ {\lambda}_1 (s)\, a_1 (\chi_1,\chi_2) +  {\lambda}^{\prime}_1 (s)\, b_1 (\chi_1,\chi_2) + \frac{\tilde{G}(s)}{{\eta}_{\circ}} \Big(   {\lambda}_1 (s) \, a_2 (\chi_1,\chi_2) +  {\lambda}^{\prime}_1 (s) \, b_2 (\chi_1,\chi_2)  \Big) \right] ,
%\frac{\tilde{G}(s)}{\eta_0}\left[a(\chi_1,\chi_2,s)\tilde{\lambda}_1(s)+ b(\chi_1,\chi_2,s)\tilde{\lambda}^\prime_1(s) \right]+\frac{U_V}{s} \quad
%\text{Moslem: we need the expression for $a$ and $b$} 
\label{eq:VEmain}
\ee 
where ${\chi}_1 = a/c$, ${\chi}_2 = b/c$, and
\begin{align}
 \Delta &= 5 {\chi_2}^5 ({\chi_1}^5 -1) + \frac{\tilde{G}(s)}{{\eta}_{\circ}} \big( 3 {\chi_2}^{10} + 2  {\chi_2}^5 {\chi_1}^5 -2 {\chi_1}^{5} -3 {\chi_2}^{5}  \big) , \nn \\
 a_1 (\chi_1,\chi_2) &= 5 {\chi_2}^5 ({\chi_1}^5 + 5 {\chi_1}^2 -6 ) , \qquad b_1 (\chi_1,\chi_2) = 5 {\chi_2}^5 (-2{\chi_1}^5 + 5 {\chi_1}^2 -3 ) , \nn \\
 a_2  (\chi_1,\chi_2) &= 3 {\chi_2}^{10} + 2 {\chi_1}^5 {\chi_2}^5 -12 {\chi_1}^5 -18 {\chi_2}^5 + 15 {\chi_2}^7 + 10 {\chi_1}^5 {\chi_2}^2  ,  \nn \\
 b_2  (\chi_1,\chi_2) &=  -6 {\chi_2}^{10} -4 {\chi_1}^5 {\chi_2}^5 -6 {\chi_1}^5 -9 {\chi_2}^5 + 15 {\chi_2}^7 + 10 {\chi_1}^5 {\chi_2}^2 . \nn
\end{align}
Using the experimentally measured velocities of the cell surface and nucleus in Eq.\eqref{eq:VEmain}, we can compute ${\tilde{G}(s)}/{\eta_{\circ}}$.  
This can then be inverted to real space to get ${G(t)}/{\eta_{\circ}}$. Note that we can only compute $G(t)/\eta_{\circ}$ and not the individual 
variables independently.  This is in line with our findings for viscous cytoplasm, where the nucleus velocity was independent of viscosity 
and only a function of geometry. 

The process of characterizing the linear viscoelastic response from experiments  can be summarize as follows:
\begin{enumerate}
\item Measure the cell surface and nucleus displacements (and, thus, velocities) with time and measure the intracellular distribution of filaments by fluorescent microscopy to approximate the outer radius of the shell in the model. 
\item Decompose the cell surface rate of displacements (velocities) in radial and angular directions into Legendre polynomials: 
$v_{r}  = \sum_{n=1}^{\infty} {\lambda}_n P_n (\cos\theta)$ and $v_{\theta} = \sum_{n=1}^{\infty} {\lambda}_{n}^{\prime} V_n (\cos\theta)$. 
\item Take Laplace transform of the surface velocity models, $\lambda(t)$ and $\lambda^\prime(t)$, the nucleus velocity $U_{\mathrm{VE}}(t)$ to get 
$\tilde{\lambda}(s),\, \tilde{\lambda}^\prime(s)$ and $\tilde{U}(s)$. 
\item Use Eq.\eqref{eq:VEmain} to compute $\tilde{G}(s)/\eta_{\circ}$, and take inverse Laplace of $\tilde{G}(s)$ to get $G(t)/\eta_{\circ}$. 
To compute the linear oscillatory functions, $G^\prime (\omega)$ and $G^{\prime\prime} (\omega)$ one can 
replace $s$ in $\tilde{G}(s)$ with $i\omega$ and use the expression $G^* (\omega)=i\omega \tilde{G}(i \omega)$ to compute 
$G^*(\omega)$. The real and imaginary parts of $G^*$ correspond to $G^\prime$ and $G^{\prime\prime}$, respectively.   
\end{enumerate}

 Thus far we have outlined how experimentally provided values of $U_{\mathrm{VE}}(t)$ and $\lambda(t)$ and $\lambda^\prime(t)$ can 
be used to compute the linear rheology of the cytoplasm. 
In the next section we use the simplest viscoelastic constitutive equation, namely as Maxwell fluid, to explore the role of geometrical parameters (such as $\chi_2$) and physical parameters (such as 
the relaxation time $\tau$) 
on the time-dependent response of $U_{\mathrm{VE}}$.

However, large number of studies show that the rheology of cytoplasm is far more complex than a Maxwell fluid \cite{hoffman2009cell}. 
The model that has emerged out of these studies suggest that the dynamic moduli are described by a sum of two power-laws \cite{hoffman2006consensus}:
\begin{subequations}
\begin{align}
& G^\prime(\omega)=A\cos (\pi \beta/2)\omega^\beta+B\cos(3\pi/8)\omega^{3/4}, \\
& G^{\prime\prime}(\omega)=A\sin (\pi \beta/2)\omega^\beta+B\sin(3\pi/8)\omega^{3/4},
\end{align}
 \label{eq:powerlaw}
 \end{subequations}
where $0.15 <\beta<0.25$ and $A$ and $B$ are constants dependent on the biological system. 
In the \emph{MATLAB Live Script} file we reproduce the results of the next section, when the constitutive equation is given by Eqs.\eqref{eq:powerlaw}.

 \subsubsection{Example: Maxwell fluid}
We model the fluid within the shell using a Maxwell constitutive equation, which is given by 
\be 
(1+ \tau \frac{\partial}{\partial t} ) {\boldsymbol{\sigma}}^{\mathrm{p}} = {\eta}_p \mathbf{D}.\label{Maxwell} 
\ee
Here $\tau=\eta_p/G_p$ is the relaxation time of the polymer phase, and $\eta_p$ and $G_p$ are the shear viscosity and modulus of the polymer phase.
Although the nucleus velocity can be computed for any time-dependent choice of functions of $\lambda (t)$ and $\lambda^\prime (t)$,
we choose the simplest case of constant velocity modes $\lambda=\lambda^\prime=1$. This choice provides an analogy
to creep and stress relaxation experiments in shear rheology. 
Taking the Laplace transform of the shell stress and the surface velocities gives 
$\tilde{G}(s)=\frac{\eta_p} {1+s\tau}$, and $\tilde{\lambda}=\tilde{\lambda}^\prime=\frac{1}{s}$. 
We can compute the nucleus velocity in Laplace space by substituting for $\tilde{G}(s)$, $\tilde{\lambda}(s)$ 
and $\tilde{\lambda}^\prime(s)$ in Eq.\eqref{eq:VEmain}. Inverting from Laplace to time space gives 
\begin{align}
&& U_{\mathrm{VE}} (t) =  f_1 ({\chi}_1 , {\chi}_2 , {\eta}^{\star}) {\lambda}_1 +  f_2 ({\chi}_1 , {\chi}_2 , {\eta}^{\star}) {\lambda}_{1}^{\prime} + ({\lambda}_1 + {\lambda}_{1}^{\prime})   f_3 ({\chi}_1 , {\chi}_2 , {\eta}^{\star}) {\mathrm{e}}^{-t/{\tau}_{\mathrm{VE}}} ,
\end{align}
where $\eta^\star=\frac{\eta_p}{\eta_\circ}$ and 
\begin{align}
&&& f_1  = \frac{1}{\Delta} \Big( 5 {{\chi}_1}^5 {{\chi}_2}^5 -30 {{\chi}_2}^5 + 25 {{\chi}_1}^2 {{\chi}_2}^5  + \nn \\
&&& \qquad\qquad\qquad\qquad + {\eta}^{\star} \left( 3 {{\chi}_2}^{10}  + 2 {{\chi}_1}^5 {{\chi}_2}^5 - 12 {{\chi}_1}^5 -18 {{\chi}_2}^5  + 15 {{\chi}_2}^7 + 10 {{\chi}_1}^5 {{\chi}_2}^2  \right)  \Big) \nn \\
&&&  f_2 = \frac{1}{\Delta} \Big( -10 {{\chi}_1}^5 {{\chi}_2}^5 -15 {{\chi}_2}^5 + 25 {{\chi}_1}^2 {{\chi}_2}^5  +  \nn \\
&&& \qquad\qquad\qquad\qquad  {\eta}^{\star} \left( -6 {{\chi}_2}^{10}  -4 {{\chi}_1}^5 {{\chi}_2}^5 - 6 {{\chi}_1}^5 -9 {{\chi}_2}^5  + 15 {{\chi}_2}^7 + 10 {{\chi}_1}^5 {{\chi}_2}^2  \right)  \Big) \nn \\ 
&&&  f_3  =  \frac{5 {\eta}^{\star} (1- {\chi}_2) (2 {{\chi}_1}^6 -2 {{\chi}_1}^5 {\chi}_2 + 3 {\chi}_1 {{\chi}_2}^5 - 3 {{\chi}_2}^6  )}{\Delta ( 1+ {\chi}_1 + {{\chi}_1}^2 + {{\chi}_1}^3 + {{\chi}_1}^4  )} \times \nn \\
&&& \qquad \times  \left( {{\chi}_1}^3 {\chi}_2 + {{\chi}_1}^3 + {{\chi}_1}^2 + {\chi}_1 +  {{\chi}_1}^2  {{\chi}_2}^2   + {\chi}_1 {{\chi}_2}^3 + {{\chi}_2}^3 + {{\chi}_2}^2 + {\chi}_2  + 2 {\chi}_1 {\chi}_2 (1+ {\chi}_1 + {\chi}_2)   \right), \nn  \\
&&& \Delta = 3 \left( 5 {{\chi}_1}^5  {{\chi}_2}^5  - 5 {{\chi}_2}^5  + {\eta}^{\star} \left( 3 {{\chi}_2}^{10} + 2 {{\chi}_1}^5 {{\chi}_2}^5  -2 {{\chi}_1}^5 -3 {{\chi}_2}^5   \right)   \right), \nn
\end{align}
with
\be 
{\tau}_{\mathrm{VE}} = \tau \, \frac{5 {{\chi}_2}^2 (1- {{\chi}_1}^5)}{5 {{\chi}_2}^2 (1- {{\chi}_1}^5) + {\eta}^{\star} \left( (1- {{\chi}_2}^5) (3 {{\chi}_2}^5  + 2 {{\chi}_1}^5)  \right)  } . \label{eq:tauVE}
\ee

\begin{figure}
    \centering
    \subfigure[]{\includegraphics[width=0.47\textwidth , height=6.5cm ]{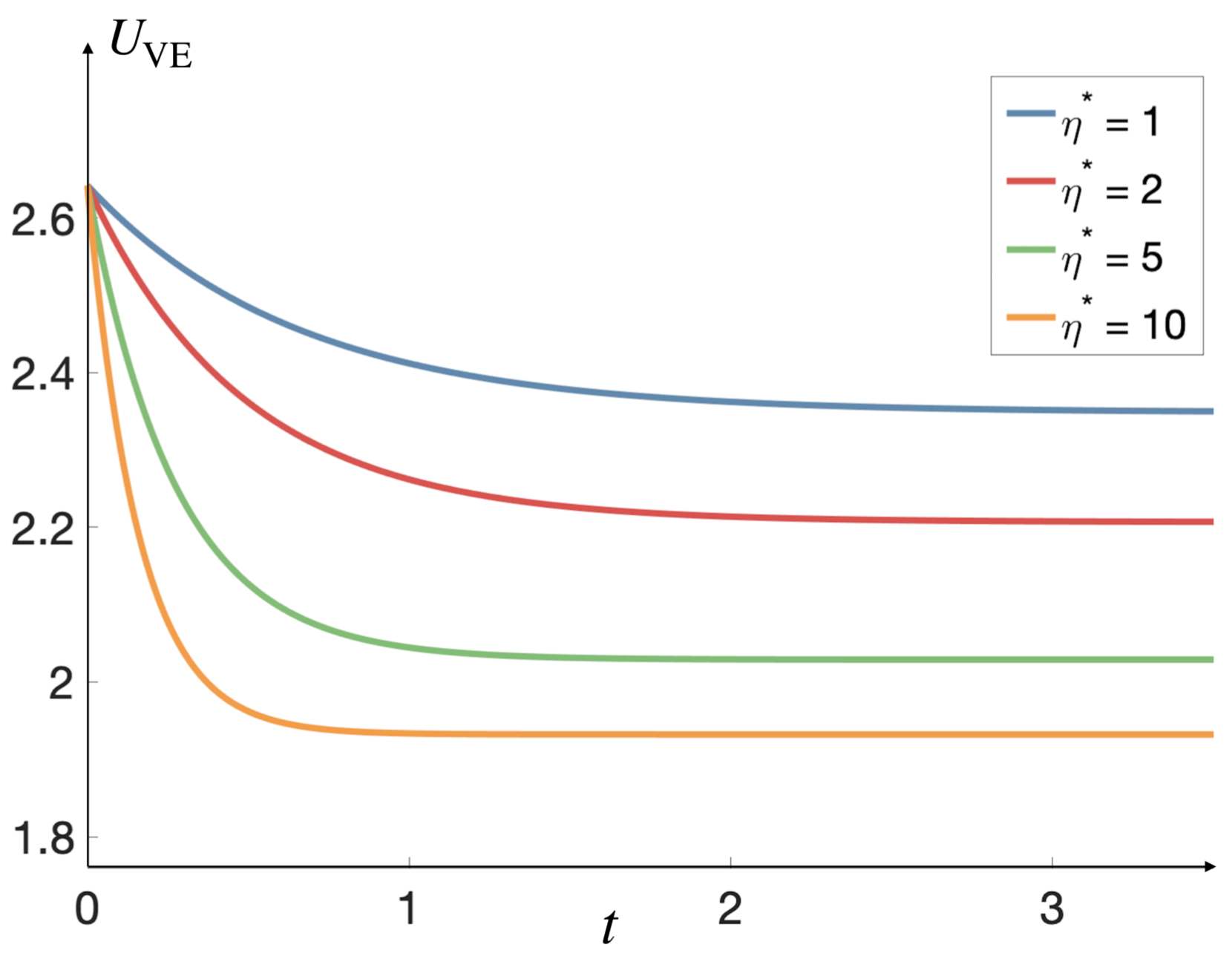}} \label{Fig3a} \quad
    \subfigure[]{\includegraphics[width=0.47\textwidth , height=6.5cm]{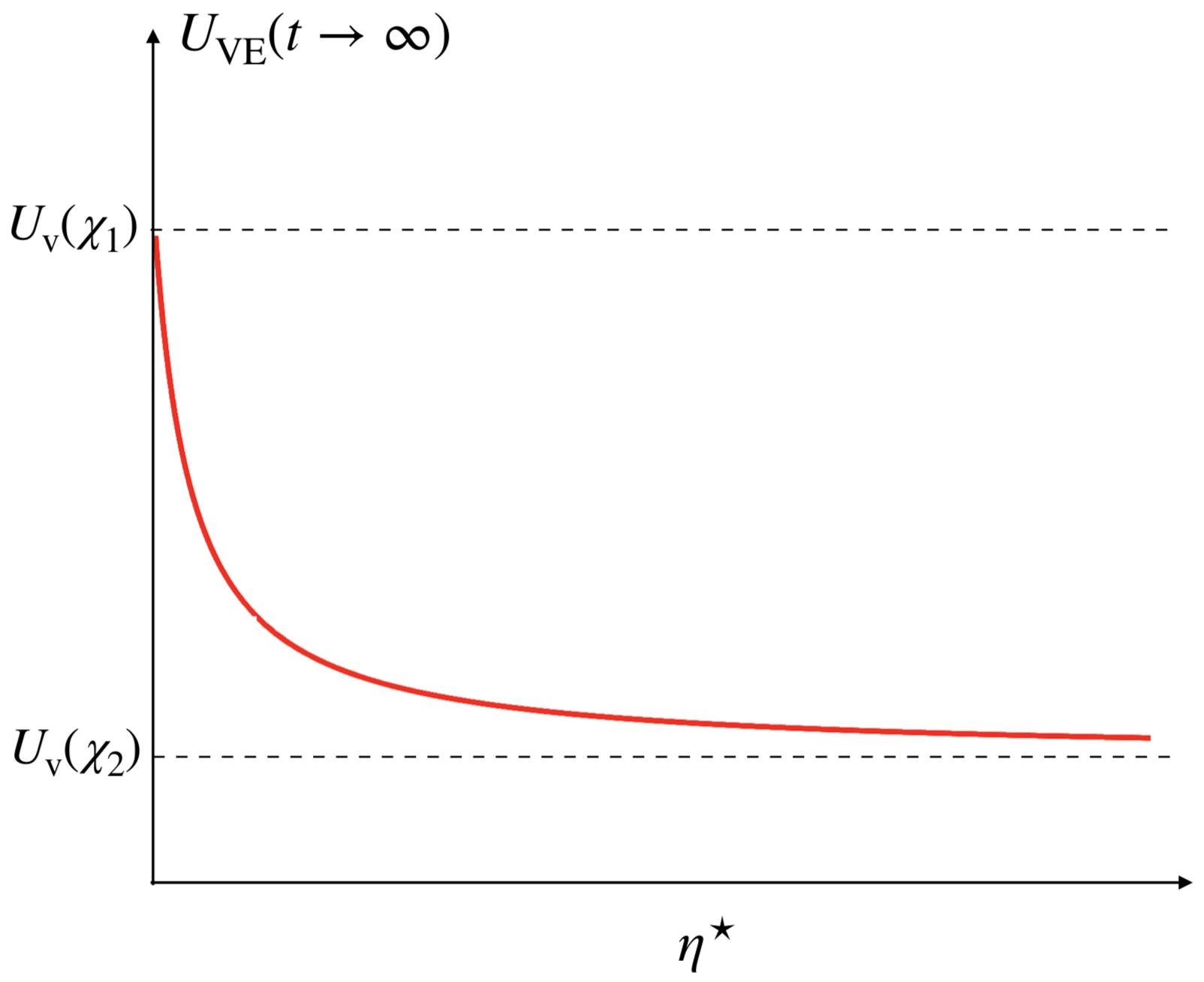}} \label{Fig3b} 
    \caption{(a)  Velocity of the nucleus as a function of time for different values of the relative viscosity, ${\eta}^{\star}$, in a cytoplasm described by Maxwell constitutive equation, when ${\chi}_1=1/3, {\chi}_2 = 2/3, {\lambda}_1 =1, {\lambda}_{1}^{\prime} =1$ and $\tau =1$. 
    (b) The long-time velocity of the nucleus moving as a function of the ratio of polymeric to solvent viscosities in Maxwell constitutive equation, ${\eta}^{\star}$. In the limit of ${\eta}^{\star} \to 0$, we recover the velocity of a nucleus with radius $r=a$, moving in a Newtonian cytoplasm, $U_{\mathrm{v}}(\chi_1)$; and when ${\eta}^{\star} \to \infty $, we recover the velocity of a nucleus with radius $r=b$, moving in a Newtonian cytoplasm, $U_{\mathrm{v}}(\chi_2)$. 
     }
    \label{fig3}
\end{figure}

Figure \ref{fig3}(a) shows the velocity of the inner sphere as a function of time for $a=1, b=2, c=3, {\lambda}_1 =1$, and $ {\lambda}_{1}^{\prime} =1$.
At $t=0$, $\boldsymbol{\sigma}_p=\mathbf{0}$ and the velocity of the nucleus is that of a solid sphere of radius $r=a$ in a Newtonian fluid with viscosity $\eta_{\circ}$. At long times the steady-state velocity of the nucleus is a function of relative viscosity, ${\eta}^{\star}$. 
The limit of $\eta^\star=0$ corresponds to negligible contribution to stress from the polymeric phase and we recover the velocity of the inner sphere for a Newtonian fluid with viscosity $\eta_{\circ}$. 
In the limit of $\eta_p \to \infty$, the outer shell behaves asymptotically as a solid sphere; thus, the nucleus velocity asymptotes to the velocity of a nucleus with radius $r=b$ in a Newtonian fluid of viscosity $\eta_{\circ}$;  see Fig.\ref{fig3}(b). 
Thus, $\eta^\star$ can be determined by the long-time (small frequency) values of the nucleus velocity and the relaxation time can be determined by its intermediate values.  

\begin{figure}
    \centering
    \begin{minipage}[t]{.29\linewidth}
    \subfigure[]{\includegraphics[width=\textwidth  ]{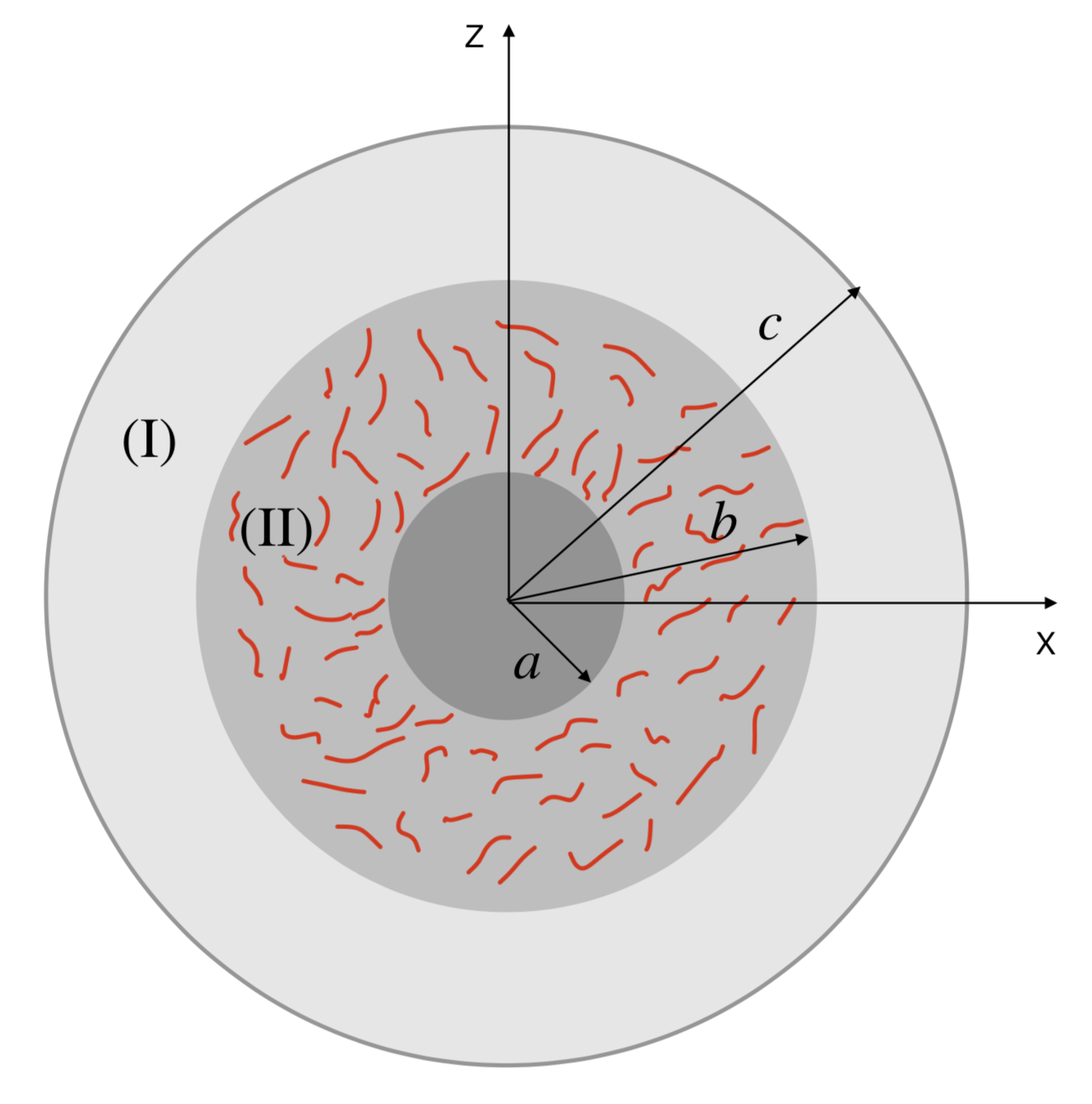}} \label{Fig4a}\\
    \subfigure[]{\includegraphics[width=\textwidth ]{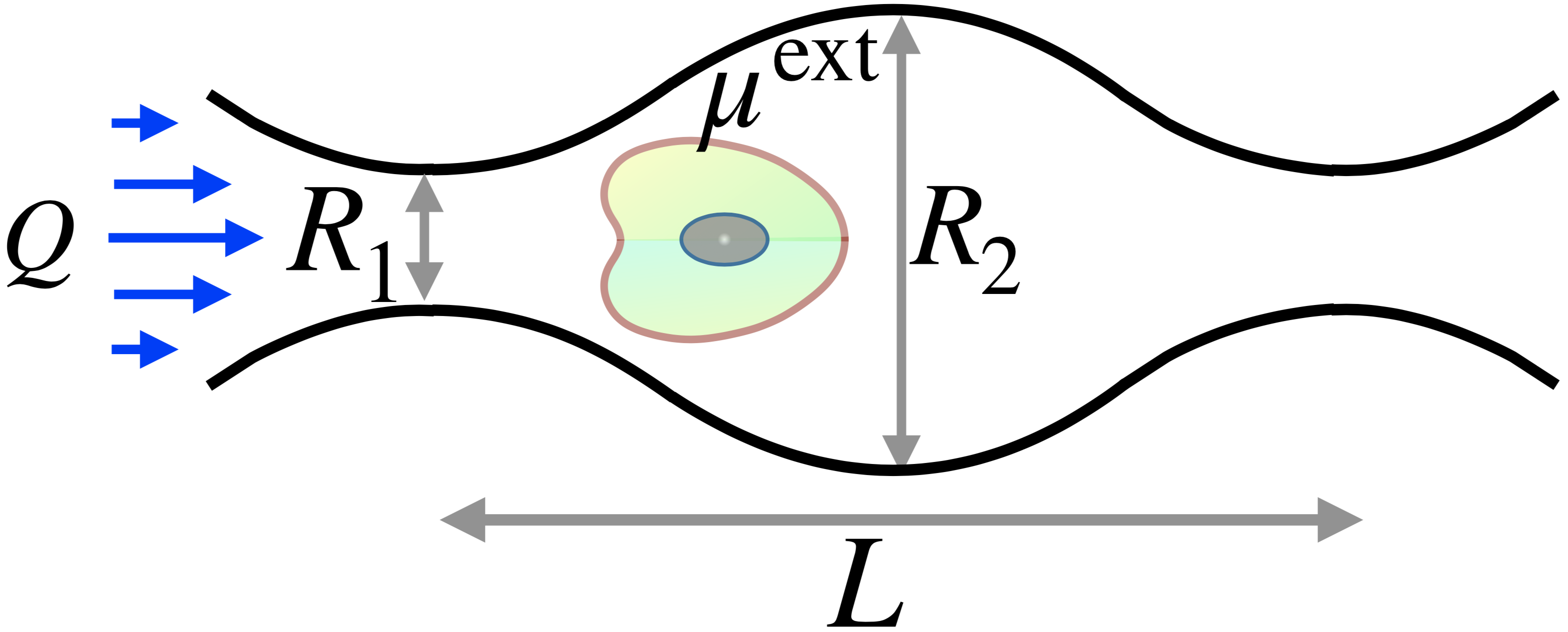}} \label{4b} 
    \end{minipage}
    \hspace{0.2in}
    \begin{minipage}[t]{.6\linewidth}
                \subfigure[]{\includegraphics[width=0.9\textwidth, height=8.05cm ]{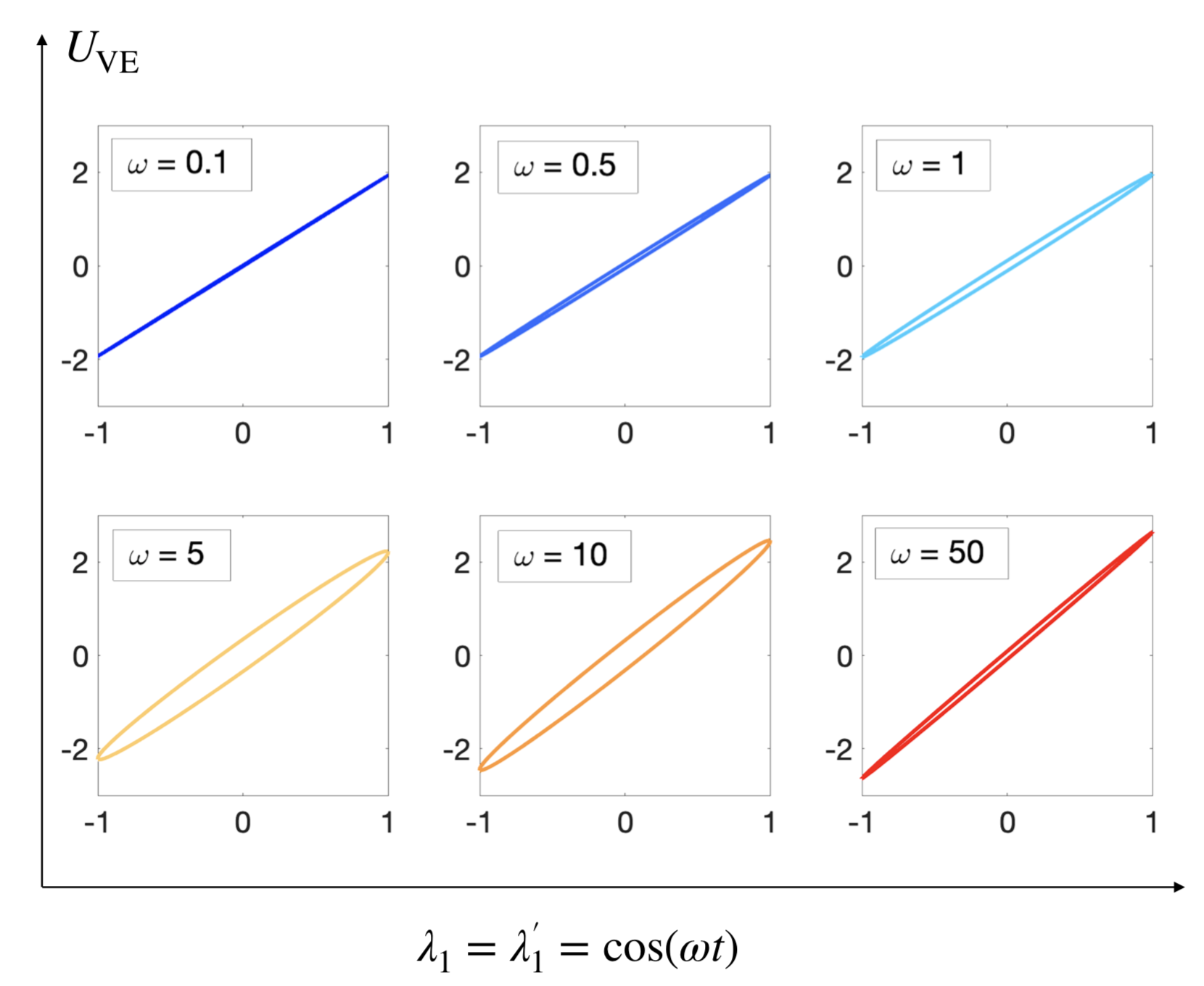}} \label{Fig4c} 
        \end{minipage}
    \caption{(a) A schematic representation of a viscoelastic shell,
    adjacent to  the nucleus of radius $r=a$ and cortex with radius $r=b$, containing freely suspended flexible filaments.
    (b) A schematic representation of a cell moving through a 
    microchannels with periodic contractions and expansions.   
    This periodicity in geometry results in periodic external hydrodynamic forces and cell deformations. 
    The frequency and amplitude of these deformations can, in principle, be controlled through changing 
    the microchannels geometry ($R_1$, $R_2$, and $L$), as well as the fluid viscosity and flow rate ($\mu^{ext}$, and $Q$). 
    (c) Velocity of the nucleus vs. the velocity of the cortex, for periodic deformation of the cortex, ${\lambda}_1 ={\lambda}_{1}^{\prime} =\cos(\omega t)$,
    at different frequencies, $\omega$, when the cytoplasm is described by Maxwell constitutive equation with relaxation time $\tau=1$, and $\eta^\star=10$.}
    \label{fig4}
\end{figure}

Next, we consider periodic deformations of the cell cortex as an analogy to linear oscillatory shear rheology. 
Unlike the case of constant values of $\lambda$ and $\lambda^\prime$, periodic deformations can be 
experimentally realized e.g.  by periodic regions of contractions and expansion; see Fig.\ref{fig4}(b) for a schematic. A wide range of frequencies and magnitude of these deformations can, 
in principle, be explored by tuning the external viscosity, flow rate and 
 the geometrical features of the microchannel.  
If we assume that the cortex (cell surface) displacement is sinusoidal with frequency $\omega$ namely $\lambda=\lambda_\circ\cos \omega t$, 
we expect the nucleus velocity to have a frequency-dependent phase-lag ($\delta$) 
in response to these surface deformations, $U_{\mathrm{VE}}(t)=U_{\mathrm{VE}}^{0}\cos (\omega t+\delta)$.
For purely viscous fluid, this phase-lag, $\delta$, is identically zero. Thus, plotting $U_{\mathrm{VE}} (t)$ vs $\lambda(t)$ and $\lambda^\prime(t)$  
will produce a line. For a purely elastic response the phase-lag will be $\delta=\pi/2$, 
which will give circles for $U_{\mathrm{VE}} (t)$ vs $\lambda(t)$ and $\lambda^\prime(t)$. 
For a linear viscoelastic fluids, $0<\delta<\pi/2$ and $U_{\mathrm{VE}} (t)$ vs $\lambda(t)$ and $\lambda^\prime(t)$ curves would be tilted ellipses.

These curves are commonly known as Lissajous curves in rheology and they provide a powerful tool for 
nonlinear rheological fingerprinting of complex fluids using 
large amplitude (nonlinear) oscillatory rheology \cite{hyun2011review}. 
Although we have limited our analysis to linear regime, $U_{\mathrm{VE}}-\lambda_n$ curves 
and the mathematical framework developed to analyze these curves can, in practice, 
be deployed to characterize the linear and nonlinear viscoelastic 
behavior of the cytoplasm over a wide range of forces and frequencies. 
As an example, Fig.\ref{fig4}(b) shows $U_{\mathrm{VE}}(t)$ vs $\lambda_1 (t) = \lambda_1^\prime(t)=\cos \omega t $ 
at different frequencies using Eq.\eqref{eq:VEmain}. 
We see that, the area within tilted ellipse 
is increased, and decreased, showing complex rheological response of the system. 
In linear shear oscillatory rheology the geometric properties of this curve, such as the area 
within the ellipse and its maxima, can be used to extract $G^\prime(\omega)$ and $G^{\prime\prime}$ \cite{hyun2011review}. 
Analogous relationships can be derived in our case using Eq.\eqref{eq:VEmain},
which relates the curve characteristics to material parameters such as $G(t)/\eta_{\circ}$. 
However, one important difference with the shear rheology is that in our case these 
relationships will also be dependent on the details of fluid flow geometry i.e. $\chi_1$ and $\chi_2$.  
We will not derive these relationships here. 

%\begin{figure}
  %\centerline{\includegraphics[width= \textwidth]{4}}% Images in 100% size
  %\caption{ (a) Schematic of viscoelastic cytoplasm floating freely around the nucleus in concenteric spherical geometry. (b) The velocity of the nucleus vs. the velocity of the cortex, for periodic deformation of the cortex, and ${\lambda}_1 ={\lambda}_{1}^{\prime} =\cos(\omega t)$ in viscoelastic cytoplasm.}
%\label{fig4}
%\end{figure}
%
%\begin{figure}
%    \centering
%    \subfigure[]{\includegraphics[width=0.4\textwidth , height=7cm ]{4a}} \label{Fig4a} \quad
%    \subfigure[]{\includegraphics[width=0.55\textwidth , height=7.5cm]{4b}} \label{Fig4b} 
%    \caption{(a)  Schematic of viscoelastic cytoplasm floating freely around the nucleus in concentric spherical geometry. 
%    (b) Velocity of the nucleus vs. the velocity of the cortex, for periodic deformation of the cortex, ${\lambda}_1 ={\lambda}_{1}^{\prime} =\cos(\omega t)$, in viscoelastic cytoplasm for different frequencies.
%     }
%    \label{fig4}
%\end{figure}
\subsubsection{Special case of \texorpdfstring{$\chi_2=1$}. } 
For the special case of ${\chi}_2 =1$, the velocity of the nucleus will be independent of $\tilde{G}(s)$, and the coefficients in the 
Eq.\eqref{eq:VEmain} are: $\Delta = 5 ({{\chi}_1}^5 -1)$, $a_1 = 5 ({{\chi}_1}^5 + 5 {{\chi}_1}^2 -6 )$, $b_1 = 5 (-2 {{\chi}_1}^5 + 5 {{\chi}_1}^2 -3 )$, $a_2 = b_2 =0$. Thus, 
\be 
\tilde{U}(s)= \frac{1}{3\Delta} \left[ {\lambda}_1 (s)\, a_1 (\chi_1) +  {\lambda}^{\prime}_1 (s)\, b_1 (\chi_1) \right].
\ee
For constant first modes ${\lambda}_1 (s) = {\lambda}^{\prime}_1 (s) = \frac{1}{s}$, this velocity becomes the velocity of nucleus in viscous medium, and we recover Eq.\eqref{U1solid}. Hence, 
structural heterogeneity within the cytoplasm must be present to observe the nontrivial nucleus velocities discussed here. This special case is physiologically improbable as the cytoskeletal organizations are almost always heterogeneously distributed. Conversely, this analysis also point to the importance of accounting for these heterogeneities in our models. 

\subsection{Porous shell}
Thus far we have described the cytoplasm mechanics, using a general linear viscoelastic constitutive equation to model the presence of 
the cytoskeletal assemblies. 
The underlying assumption in this coarse-graining is that the net force on each filament, and the filament phase as a whole, is identically 
zero. The force moment on the filaments, however, is non-zero giving rise to an extra stress in the continuum limit, which is modeled through the appropriate constitutive equations. 

The force-free condition for filaments is valid as long as they are freely suspended, with no constraint in their motion. 
In cytoskeletal assemblies the filaments are often crosslinked together or to other structures via motor proteins and passive crosslinkers. In other words, the filaments form networks, rather than suspensions. For example, microtubules typically nucleate from microtubule organizing centers, which are attached to the nuclear envelope; see Fig.\ref{fig1}(a). Thus, microtubule array and the nucleus move together.  
Because of the constrains in their motion, the filaments (microtubules) experience a net drag force 
due to the relative motion of the cytoplasmic fluid with respect to the network (array). 

On the continuum scale, this force can be modeled through a body force acting on the network that scales 
linearly with the velocity difference between the network and the fluid, 
$\mathbf{f}^\text{porous}(\mathbf{x}) \sim \left(\mathbf{v}(\mathbf{x})-\mathbf{U}\right)$.  
In other words, the filament network around the nucleus behaves as a porous medium. 
In the first step, we neglect the elastic properties of the network and assume the network is made of rigid filaments with uniform orientation and density. 
Assuming that the cytoplasmic fluid permeating  the network is Newtonian,  the fluid flow within the network can be modeled using 
Brinkman equation:
\begin{align}
&a<r<b:&  &\eta_{\circ} {\nabla}^2 \mathbf{v} -  \frac{\eta_{\circ}}{\kappa} \mathbf{v}   - \nabla p =\mathbf{0}& &\mathrm{and}&   &\nabla \cdot \mathbf{v} =0,&
\label{eq:Brinkman} 
\end{align} 
where $\kappa = \frac{1}{{\alpha}^2}$ is the permeability of the medium. 
Note that in Eq.\eqref{eq:Brinkman} the velocity field, $\mathbf{v}(\mathbf{x})$, 
is defined in the coordinates that moves with the nucleus, such that $\mathbf{v}(\mathbf{x})=\mathbf{0}$ 
when $\mathbf{x} \in S_a$.

Here we provide a simple scaling analysis of fluid permeation in cytoskeletal networks. 
The volume fraction of these networks within the cytoplasm is typically small, $\phi \ll 1$.  
Sangani \& Acrivos \cite{sangani1982slow} give an asymptotic expression for permeability: 
\begin{equation}
\frac{\kappa}{a_f^2}=\frac{1}{8\phi} \left(-\ln \phi -1.48\right),
\label{eq:kappa}
\end{equation}
where $a_f$ is the radius of the fiber. 
From this expression we see that since $\phi\ll 1$, the fluid permeability 
is large in the length-scale of the fiber radius, $\kappa/a_f^2 \gg 1$. 
For microtubules, which have the largest radius among cytoskeletal filaments, $a_f \approx12 \text{nm}$. 
As a result, the ratio of nucleus to fiber radius is typically $\mathcal{O}(10^3)$. 
Rescaling this expression to the nucleus radius --which is the scale of interest here-- gives 
$\frac{\kappa}{a^2}= \left(\frac{a_f}{a}\right)^2\frac{1}{8\phi} \left(-\ln \phi -1.48\right)$. 
Even for very small volume fractions we have the likely  condition of $\phi (a/a_f)^2 \gg 1$, 
which makes $\kappa/a^2 \le 1$ resulting in substantial
reduction in fluid permeability by the network in the nucleus (and the cell) length-scale. In other words, 
the network can have large effects on the fluid flows 
and the nucleus dynamics even when $\phi \ll 1$. 

\begin{figure}
    \centering
    \subfigure[]{\includegraphics[width=0.37\textwidth , height=7.3cm ]{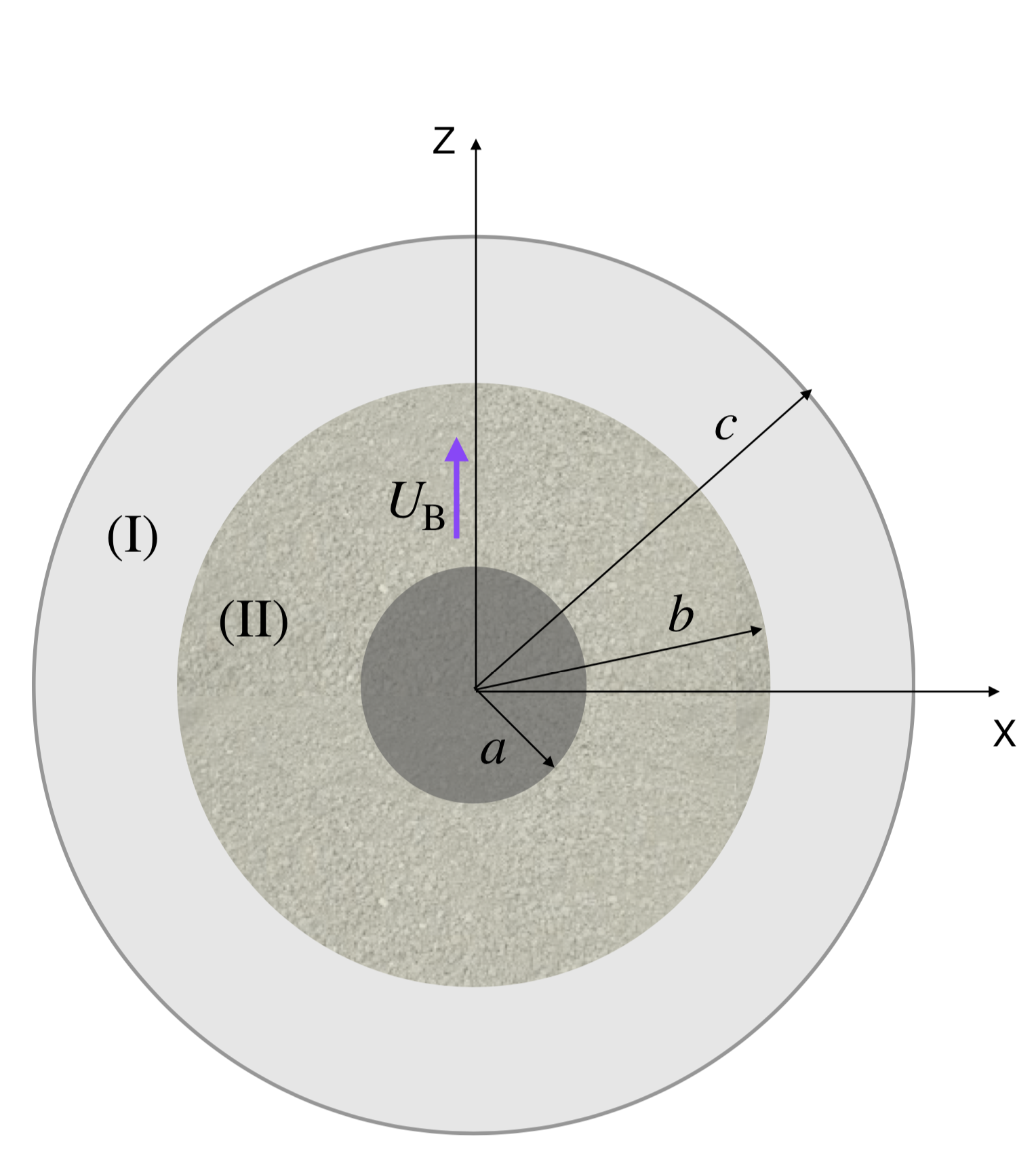}} \label{Fig5a} \quad
    \subfigure[]{\includegraphics[width=0.53\textwidth , height=7.8cm]{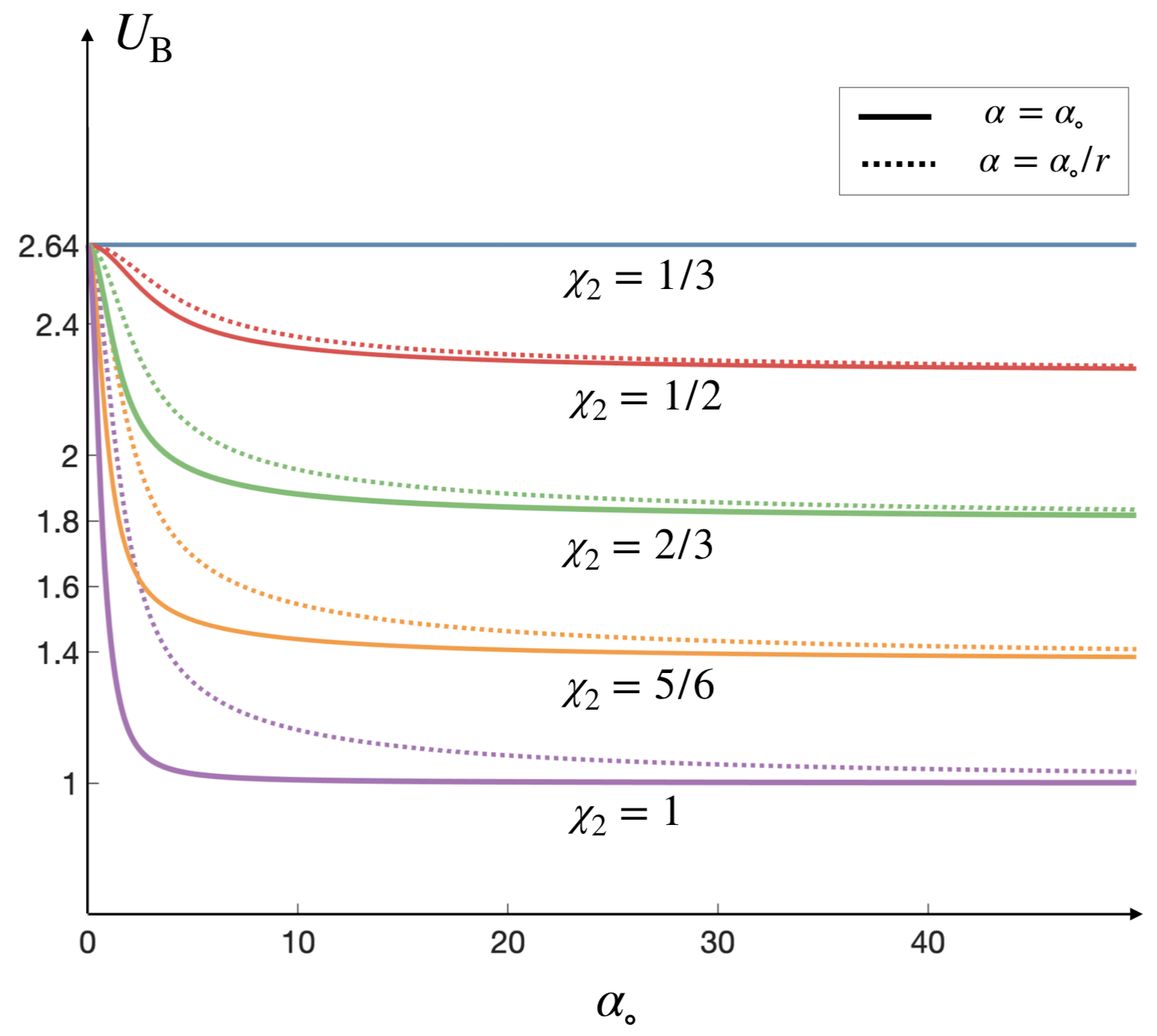}} \label{Fig5b} 
    \caption{(a)  A schematic representation of a porous shell of outer radius $r=b$, surrounding the nucleus of radius $r=a$. The shell is composed of filaments anchored to the nucleus and crosslinked together.
    (b) Velocity of the nucleus as a function of $\alpha_0=1/\sqrt{\kappa}$, for different values of ${\chi}_2 = \frac{b}{c}$.The dashed line shows the same results for spatially variant permeability $\alpha=\alpha_0/r$ where $r$ is the distance from the center of the nucleus. As expected, at a given $\alpha_0$ 
    the average $\alpha$ of the radially decaying $\alpha$ is less than 
    the constant permeability case, leading to larger values for the nucleus velocity.
    The limiting values at $\alpha_0 \to 0$ and $\alpha_0 \to \infty$ are identical for both constant and spatially variant permeabilities, and correspond to the nucleus velocity in a Newtonian cytoplasm with radius $r=a$ and $r=b$, \textit{i.e.} $U_{\mathrm{v}}(\chi_1)$ and $U_{\mathrm{v}}(\chi_2)$, respectively.}
    \label{fig5}
\end{figure}

Here we consider a porous cytoplasm with a rigid nucleus in a concentric spherical geometry; see Fig.\ref{fig5}(a). Since the flow is axisymmetric, the velocity field in spherical coordinates would be 
$\mathbf{v}(r,\theta )= (v_r , v_{\theta})$, with $\theta\in[0,\pi]$. 
The BCs and the force-free condition on the nucleus are given by
\be
\begin{aligned}
r=a:&  && {\mathbf{v}}^{(\mathrm{II})} =\mathbf{0}, && \int_{S_a} \left(\mathbf{\sigma} \cdot\mathbf{n}\right)\cdot\hat{\mathbf{z}}=0. \\ 
r=b:&  && {\mathbf{v}}^{\mathrm{(I)}} = {\mathbf{v}}^{\mathrm{(II)}} &&   
{\sigma}_{r\theta}^{(\mathrm{I})} = {\sigma}_{r\theta}^{(\mathrm{II})}, && {p}^{(\mathrm{I})} =  {p}^{(\mathrm{II})},&  \\
r=c:&  && v_{r}^{(\mathrm{I})} =  -U_{\mathrm{B}} \cos \theta +\sum_{n=1}^{\infty} {\lambda}_n P_n (\cos\theta), &&  
v_{\theta}^{(\mathrm{I})} =  -U_{\mathrm{B}}\sin \theta +\sum_{n=1}^{\infty} {\lambda}_{n}^{\prime} V_n (\cos\theta), && 
\end{aligned}  \label{eq:BrBCs}
\ee
where $V_{n} (\cos\theta) = \frac{2 \sin\theta}{n(n+1)} \frac{\mathrm{d} P_n (\cos\theta)}{\mathrm{d} (\cos\theta)} $, 
and $P_n (\cos\theta)$ are Legendre polynomials and the subscript $\mathrm{B}$ of $U$ denotes \textit{Brinkman}.
We apply these BCs using axisymmetric stream functions of Stokes and Brinkman equations 
for domains $(\mathrm{I})$ and $(\mathrm{II})$, respectively; see Fig.\ref{fig5}(a). The velocity components for mode $n=2$ of Brinkman domain are:
\be
\begin{aligned}
 {v}_{r}^{(\mathrm{II})} &=& - \left( {A}^{\prime}_{2} + \frac{{B}^{\prime}_{2}}{r^3} + 
 \frac{{C}^{\prime}_{2}}{r^2} y_2 (\alpha r)  + \frac{{D}^{\prime}_{2}}{r^2} y_{-2} (\alpha r) \right) \cos\theta ,  \\
 {v}_{\theta}^{(\mathrm{II})} &= &\left(  {A}^{\prime}_{2} -\frac{1}{2} \frac{{B}^{\prime}_{2}}{r^3} + 
 \frac{{C}^{\prime}_{2}}{2 r} y_2^{\prime} (\alpha r)  + \frac{{D}^{\prime}_{2}}{2 r} y_{-2}^{\prime} (\alpha r)   \right) \sin\theta,   
\end{aligned}
\ee
where $y_2 (\alpha r) = \alpha \cosh (\alpha r) - \frac{1}{r} \sinh (\alpha r)$, 
and $ y_{-2} (\alpha r) =  \alpha \sin (\alpha r) - \frac{1}{r} \cosh (\alpha r) $. For general $n$, $y_n (\alpha r)$ and $y_{-n} (\alpha r)$ are proportional to modified Bessel function of first and second kind, respectively; see Appendix~\ref{appA} for details. 

The fluid adjacent to the shell (region $(\mathrm{I})$) is modeled as a Stokesian fluid:
\be
\begin{aligned}
{v}_{r}^{(\mathrm{I})}&=& - \left( {A}_2 + \frac{B_2}{r^3} + C_2 r^2 + \frac{D_2}{r}  \right) \cos\theta ,  \\
{v}_{\theta}^{(\mathrm{I})} &=& \left( A_2 -\frac{1}{2} \frac{B_2}{r^3} + 2 C_2 r^2 +\frac{1}{2} \frac{D_2}{r}  \right) \sin\theta .   
\end{aligned}
\ee
%\begin{figure}
  %\centerline{\includegraphics[width= \textwidth]{5}}% Images in 100% size
  %\caption{ (a) Schematic of porous cytoplasm consisting of constrained filaments anchored to the nucleus in a concenteric spherical geometry.
  %(b) Velocity of the nucleus as a function of (inverse) permeability, for different values of ${\chi}_2 = \frac{b}{c}$; dashed line is for variable permeability. 
%  }
%\label{fig5}
%\end{figure} 

The force on the nucleus is expressed as 
\begin{eqnarray}
F &=& 2\pi b^2 \int_{0}^{\pi} \left[ {\sigma}_{rr} \sin\theta - {\sigma}_{r\theta} \cos\theta \right] {\Big|}_{r=b} \sin\theta \, \mathrm{d}\theta \nn \\
  &=&  6 \pi \eta \, U_{\mathrm{B}} b \, \, (\frac{2}{3} D_2 \sqrt{\kappa}/b ). 
\end{eqnarray}
Imposing force-free conditions, $F=0$, sets $D_2 =0$. The remaining unknown coefficients are determined by imposing BCs in Eqs.\eqref{eq:BrBCs}.
The resulting nucleus velocity, $U_{\mathrm{B}}$, is, then, calculated analytically as a function of permeability, geometrical 
parameters ${\chi}_1 = \frac{a}{c}$, and ${\chi}_2 = \frac{b}{c}$, and the first surface velocity modes ${\lambda}_1$, and ${\lambda}_{1}^{\prime}$. 
The final expression for $U_{\mathrm{B}}$ runs a few pages long and is not produced here, but is given in the \textit{MATLAB Live Script} file.

Figure \ref{fig5}(b) shows $U_{\mathrm{B}}$ vs $\alpha=\sqrt{1/\kappa}$, for different values of shell thickness ($\chi_2$), when ${\lambda}_1 = {\lambda}_1^{\prime}=1$, and $\chi_1=\frac{1}{3}$. 
 In the limit of $\alpha \to 0$, the fluid permeates the shell without any resistance and $U_{\mathrm{B}}$ asymptotes to the velocity of the Newtonian fluid of $\chi=\chi_1$. 
 On the other hand, in the limit of $\alpha \to \infty$, the fluid cannot permeate the shell and 
 $U_{\mathrm{B}}$ asymptotes to the velocity of the Newtonian fluid of $\chi=\chi_2$. 

\subsubsection{Spatially varying permeability} 
As discussed earlier, the cytoskeletal structures are heterogeneously distributed in space. 
As an example, microtubules nucleate from microtubule organizing centers and emanate radially towards the cell boundary; see Fig.\ref{fig1}(a). 
As a result, their number density decays as they grow outwards. Assuming the same length for all microtubules, the volume fraction (or number density)
decays as $\phi=\phi_\circ/r^2$. Considering $\kappa^{-1} \sim \phi$ when $\phi \ll 1$ (see Eq.~\eqref{eq:kappa}), 
we get $\kappa=\kappa_\circ r^2$ which modifies Brinkman equation to
%we get $\alpha^2=\alpha_{\circ}^2/r^2$. 
\be
\eta_\circ \nabla^2\mathbf{v}-\frac{\eta_\circ}{\kappa_\circ r^2} \mathbf{v} -\nabla p=\mathbf{0}.
\ee
In this particular case, the modified Brinkman equation allows simple analytical solutions, which are outlined in Appendix~\ref{appA}. These solutions are subjected to the same BCs
as the constant permeability given in Eq.\eqref{eq:BrBCs} to compute the nucleus velocity
as a function of $\lambda,\lambda^\prime,\chi_1,\chi_2$ and $\alpha_\circ$. 

Figure \ref{fig5}(b) compares the results of $U_{\mathrm{B}}$  as a function of ${\alpha}_{\circ}=1/\sqrt{\kappa_\circ}$, when $\alpha=\alpha_{\circ}/r$ with 
the results of constant $\alpha$ of the same value as $\alpha_{\circ}$ when $\chi_1=1/3$ and $\chi_2=2/3$ and $\lambda_1={\lambda}_1^\prime=1$. 
As expected, the values of $\alpha$ corresponding to the same value of $U_{\mathrm{B}}$ is larger for radially decaying $\alpha$ (increasing permeability).  

\subsection{Poroelastic shell}
Finally, we consider the case where the surrounding shell, $a<r<b$, is a two-phase mixture composed of a deformable 
viscoelastic network of filaments and a Newtonian  fluid permeating it. 
The momentum and mass balance equations of these two phases are given by 
\begin{align}
&\nabla \cdot {\boldsymbol{\sigma}}_f+ \frac{\eta_{\circ}}{\kappa} ({\mathbf{v}}_n - {\mathbf{v}}_f) =\mathbf{0};  
&  &\nabla \cdot \mathbf{v}_f=0 \\
&\nabla \cdot {\boldsymbol{\sigma}}_n + \frac{\eta_{\circ}}{\kappa} ({\mathbf{v}}_f - {\mathbf{v}}_n) =\mathbf{0},  &
 &\frac{\partial \phi}{\partial t}+\nabla \cdot \left( \phi {\mathbf{v}}_n\right) ={0},
 \label{eq:PE}
\end{align}
where ${\boldsymbol{\sigma}}_f=\eta_{\circ} \left(\nabla\mathbf{v}_f+\nabla\mathbf{v}_f^T\right)-p_f\mathbf{I}$ 
and ${\boldsymbol{\sigma}}_n$ are fluid and network total stress tensors, each averaged over their associated phases.  
The constitutive equation for a linear viscoelastic and isotropic network is given by 
\begin{eqnarray}
&&  {\boldsymbol{\sigma}}_n =  {\int}_0^t \left[ G(t-t^\prime)  \left( \nabla {\mathbf{v}}_n (t^\prime)+ \nabla {\mathbf{v}}_n^{T} (t^\prime) \right) +
K(t-t^\prime)\left(\nabla \cdot \mathbf{v}_n(t^\prime)\right)\mathbf{I} \right] \mathrm{d} t^\prime, 
\label{eq:PECE}
\end{eqnarray} 
where, $G$ and $K$ are the time-dependent shear and bulk moduli of the network, and $\mathbf{u}_n$ and 
${\mathbf{v}}_n  = \frac{\mathrm{d} {\mathbf{u}}_n}{\mathrm{d} t} $ are the displacement and the rate of displacement fields of the network phase.

Since the permeability is a function of 
the local volume fraction of filaments (in the limit of $\phi\ll 1$, $\kappa^{-1} \sim \phi$), the mass and momentum are coupled, leading to a 
system of nonlinear partial differential equations. 
These equations, subject to the appropriate BCs, can be solved numerically. But finding analytical solutions for them seems out of reach.
For simplicity, and for analytical tractability, we assume that the network is incompressible:  \textit{i.e.} 
its volume fraction remains uniform in space and time: $\frac{D\phi}{Dt}=\partial \phi/\partial t +\mathbf{v}_n\cdot \nabla \phi=0$.
Applying these assumptions reduces the network mass balance to $\nabla \cdot \mathbf{v}_n=0$. 
This constraint can only be imposed, without  overdetermining of our system, when $K\to \infty$ (see
 Eq.\eqref{eq:PECE}). 
In this limit, in the exact manner as Navier-Stokes equations, we would have $K(\nabla \cdot \mathbf{v}_n)=p_n$, where $p_n$ is the pressure in the network phase and a Lagrange multiplier to impose incompressibility of the network phase.    
Applying these considerations and taking Laplace transform  of Eq.\eqref{eq:PECE} yields    
\begin{equation}
 \tilde{\boldsymbol{\sigma}}_n =  \tilde{G}(s)  \left( \nabla \tilde{\mathbf{v}}_n + \nabla \tilde{\mathbf{v}}_n^{T}  \right)+\tilde{p}_n\mathbf{I}.
\label{eq:PEC2}
\end{equation} 

Next, we take Laplace transform of Eq.\eqref{eq:PE} and substitute for $\tilde{\boldsymbol{\sigma}}_n$ using 
Eq.\eqref{eq:PEC2}. The resulting equations are 
\begin{eqnarray}
&& \eta_{\circ} {\nabla}^2 {\tilde{\mathbf{v}}}_f -  \nabla {\tilde{p}}_f + \frac{\eta_{\circ}}{\kappa} ( {\tilde{\mathbf{v}}}_n - {\tilde{\mathbf{v}}}_f ) =\mathbf{0},  
\label{eq:PElaplace1} \\
&& \tilde{G}(s) {\nabla}^2 {\tilde{\mathbf{v}}}_n + \nabla {\tilde{p}}_n + \frac{\eta_{\circ}}{\kappa} ( {\tilde{\mathbf{v}}}_f - {\tilde{\mathbf{v}}}_n ) =\mathbf{0}.
\label{eq:PElaplace2}
\end{eqnarray}

Summation  and subtraction of Eqs.\eqref{eq:PElaplace1} and \eqref{eq:PElaplace2}  lead to
\begin{eqnarray}
&& \eta_{\circ} {\nabla}^2  {\tilde{\mathbf{v}}}^{+}  - \nabla {\tilde{p}}^+=\mathbf{0}, \\
&& \eta_{\circ} {\nabla}^2  {\tilde{\mathbf{v}}}^{-}  - \nabla {\tilde{p}}^- - \frac{\eta_{\circ}}{\tilde{\kappa}} {\tilde{\mathbf{v}}}^{-}  =\mathbf{0},
\label{eq:PEpm}
\end{eqnarray} 
where ${\tilde{p}}^{+}={\tilde{p}}_f - {\tilde{p}}_n$, ${\tilde{p}}^{-} = {\tilde{p}}_f + \tau(s)  {\tilde{p}}_n$, and
\begin{align*}
&  \tau(s)=\frac{\eta_{\circ}}{\tilde{G}(s)}, & &{\tilde{\mathbf{v}}}^{+}= {\tilde{\mathbf{v}}}_f + \frac{\tilde{\mathbf{v}}_n}{\tau(s)}, & 
&{\tilde{\mathbf{v}}}^{-} =  {{\tilde{\mathbf{v}}}_f -  {\tilde{\mathbf{v}}}_n }, &
%&{\tilde{p}}^{+} ={\tilde{p}}_f + {\tilde{p}}_n, & &{\tilde{p}}^{-} = {\tilde{p}}_f - \tau(s)\tilde{p}_n, &
 & { \frac{1}{\tilde{\kappa}} = \frac{1}{\kappa} \left(1+ \tau(s) \right)}. 
 \end{align*}

Note that ${\tilde{\mathbf{v}}}^{+} $ and ${\tilde{\mathbf{v}}}^{-} $ satisfy Stokes and Brinkman equations, respectively, 
for which we have analytical expressions for the velocity and stress fields. 
The BCs for the nucleus at $r=a$ and $r=c$ are identical to those 
provided for Brinkman, viscoelastic and Newtonian fluid cases. 
The boundary conditions at the interface, $r=b$, are the continuity of fluid velocity, 
tangential stress and pressure (corresponding to continuity of force across the boundary) for the fluid phase, 
and zero pressure and tangential stress for the network phase (zero force on the outer network boundary):
\be
\begin{aligned}
&r=b: \qquad & &\tilde{\mathbf{v}}^{\mathrm{(I)}} = {\tilde{\mathbf{v}}}_{f}^{\mathrm{(II)}}, \qquad
&& \tilde{\sigma}_{r\theta}^{\mathrm{(I)}} =\tilde{\sigma}_{f, r\theta}^{ \mathrm{(II)}}, 
\qquad && \tilde{p}^{\mathrm{(I)}}= {\tilde{p}}_{f}^{ \mathrm{(II)}} ,       
\qquad     &&    \tilde{\sigma}_{n, r\theta}^{ \mathrm{(II)}} =0, \qquad      
&&   {\tilde{p}}_{n}^{ \mathrm{(II)}} =0.
\label{eq:BC3}
\end{aligned} 
\ee
These BCs can be written in terms of $\tilde{\mathbf{v}}^{+}$ and  $\tilde{\mathbf{v}}^{-}$, 
by using  $\tilde{\mathbf{v}}_f=\frac{\tau \tilde{\mathbf{v}}^++\tilde{\mathbf{v}}^-}{\tau+1}$ and
$\tilde{\mathbf{v}}_n=\frac{\tau \left(\tilde{\mathbf{v}}^+-\tilde{\mathbf{v}}^-\right)}{\tau+1}$. 
The modified equations are 
\begin{subequations}
\begin{align}
&r=a: & &\tilde{\mathbf{v}}^{+}=\mathbf{0}, & &\tilde{\mathbf{v}}^{-}=\mathbf{0}.\\
&r=b: & &\tilde{\mathbf{v}}^{\mathrm(I)}=\frac{\tau \tilde{\mathbf{v}}^++\tilde{\mathbf{v}}^-}{\tau+1},  &
&\tilde{\sigma}^{(\mathrm{I})}_{r\theta}=\tilde{\sigma}^{+}_{r\theta}, &
&\tilde{p}^{(\mathrm{I})}=\tilde{p}^{+}, &
&\tilde{\sigma}^{+}_{r\theta}-\tilde{\sigma}^{-}_{r\theta}=0, &  &\tilde{p}^{+}-\tilde{p}^{-}=0.
\end{align}
\label{eq:BCspm}
\end{subequations}

We, then, use the stream function solution to Stokes and Brinkman flows 
in domains $\mathrm{I}$ and $\mathrm{II}$ and apply BCs in 
Eqs.\eqref{eq:BC1} and \eqref{eq:BCspm} to analytically determine $U_\text{PE}(s)$ 
as a function $\chi_1$, $\chi_2$, $\lambda_1$, $\lambda^\prime_1$, 
$\kappa$, $\eta_{\circ}$ and $\tau(s)$. Similar to the case of porous media, 
the final expression for $U_\text{PE}$ is very lengthy and is only given 
in the \textit{MATLAB Live Script} file. 

We can use the computed expression for $U_\text{PE}$ and follow the same four steps outlined in \textit{viscoelastic cytoplasm} section 
to compute $\kappa$ and $\eta_{\circ}/G(t)=\tau(t)$. Similarly we can compute $G^\prime(\omega)/\eta_{\circ}$ and $G^{\prime\prime}(\omega)/\eta_{\circ}$ using
$G^\star(\omega)=G^\prime(\omega)+iG^{\prime\prime}(\omega)=i\omega \tilde{G}(i\omega)$.   
Below we discuss the results for the simplest model of a network namely linear isotropic elastic model.
We present the variations of $U_{\mathrm{PE}}$ as a function of geometric parameters, including $\chi_2$, and physical parameters , including $\kappa$.
Our aim for choosing this simple, and physically inaccurate, representation of the cytoskeleton mechanics is to highlight 
the effect of simple geometrical and physical parameters such as 
$\chi_2, \kappa$ and $\tau$ on the time-dependent response of the nucleus. 
%We have included a \textit{MATLAB Live Script} that details all the analytical calculations and the 
%final expression for in this study. 
However, in the \emph{MATLAB Live Script} file we reproduce the results presented in the next section for a more physically accurate choice 
of constitutive equation of cytoskeletal assemblies given by Eq.\eqref{eq:powerlaw}.
\subsubsection{Example: linear elastic network}
We model the network mechanics using a  linear isotropic, elastic model, where $\tau=\eta_{\circ}/G$ is a constant. 
%\begin{figure}
  %\centerline{\includegraphics[width= \textwidth]{6}}% Images in 100% size
  %\caption{(a) Velocity of the nucleus as a function of time for different values of permeabilities, ${\alpha}_{\circ}$, and fixed value of elastic relaxation time, $\tau =1$. At $t=0$, this velocity is velocity of the nucleus of radius $r=a$, and at long times it is the velocity of Brinkman fluid with permeability ${\alpha}_{\circ}$. (b) Velocity of the spherical poroelastic shell as a function of time for different values of elastic relaxation time $\tau$, and fixed value of permeability ${\alpha}_{\circ} = 1$.  Here we assume ${\chi}_1 = 1/3$, ${\chi}_2 = 2/3$, and ${\lambda}_1 = {\lambda}_{1}^{\prime} =1$. }
%\label{fig6}
%\end{figure}
\begin{figure}
    \centering
    \subfigure[]{\includegraphics[width=0.48\textwidth , height=6.2cm ]{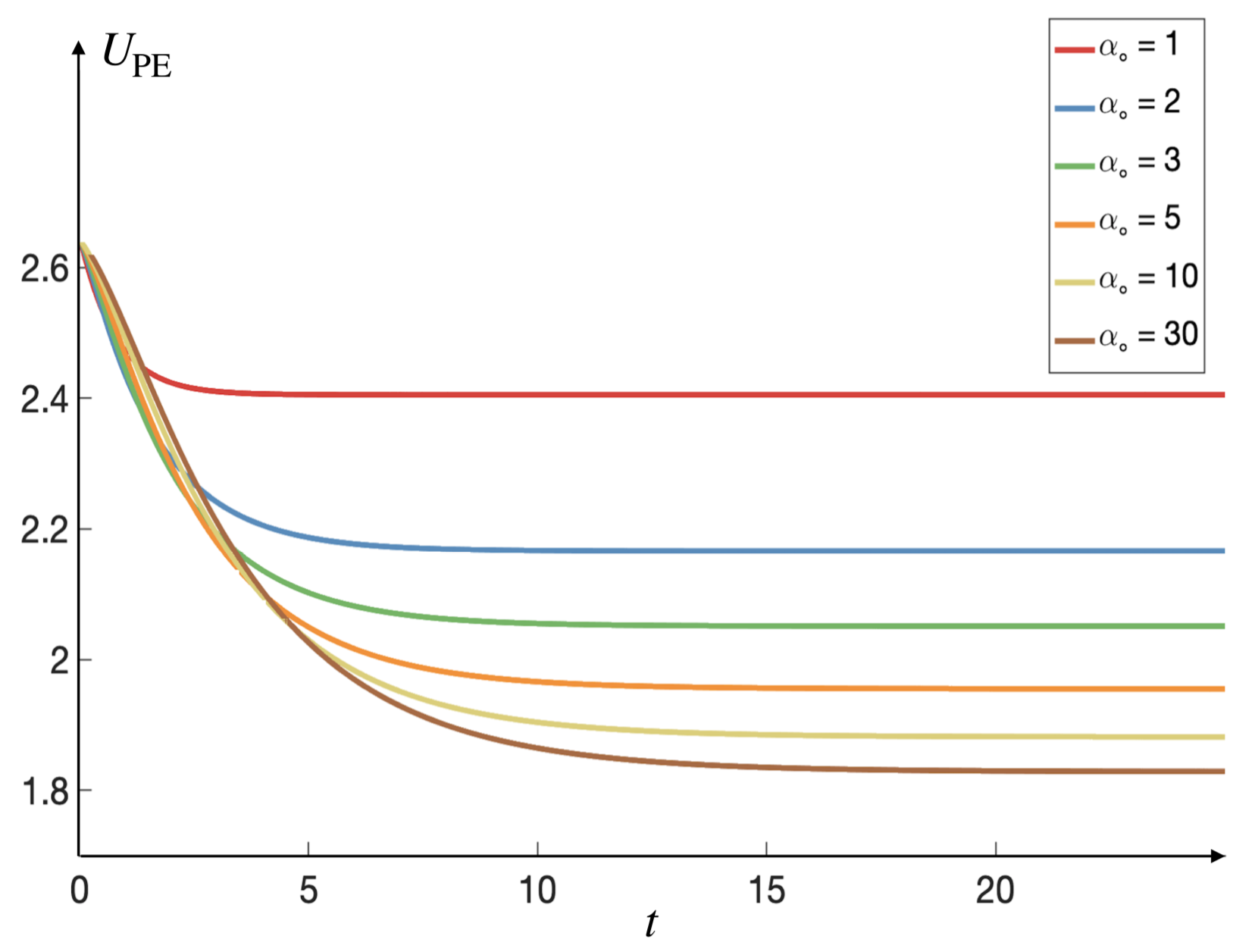}} \label{Fig6a} \quad
    \subfigure[]{\includegraphics[width=0.48\textwidth , height=6.2cm]{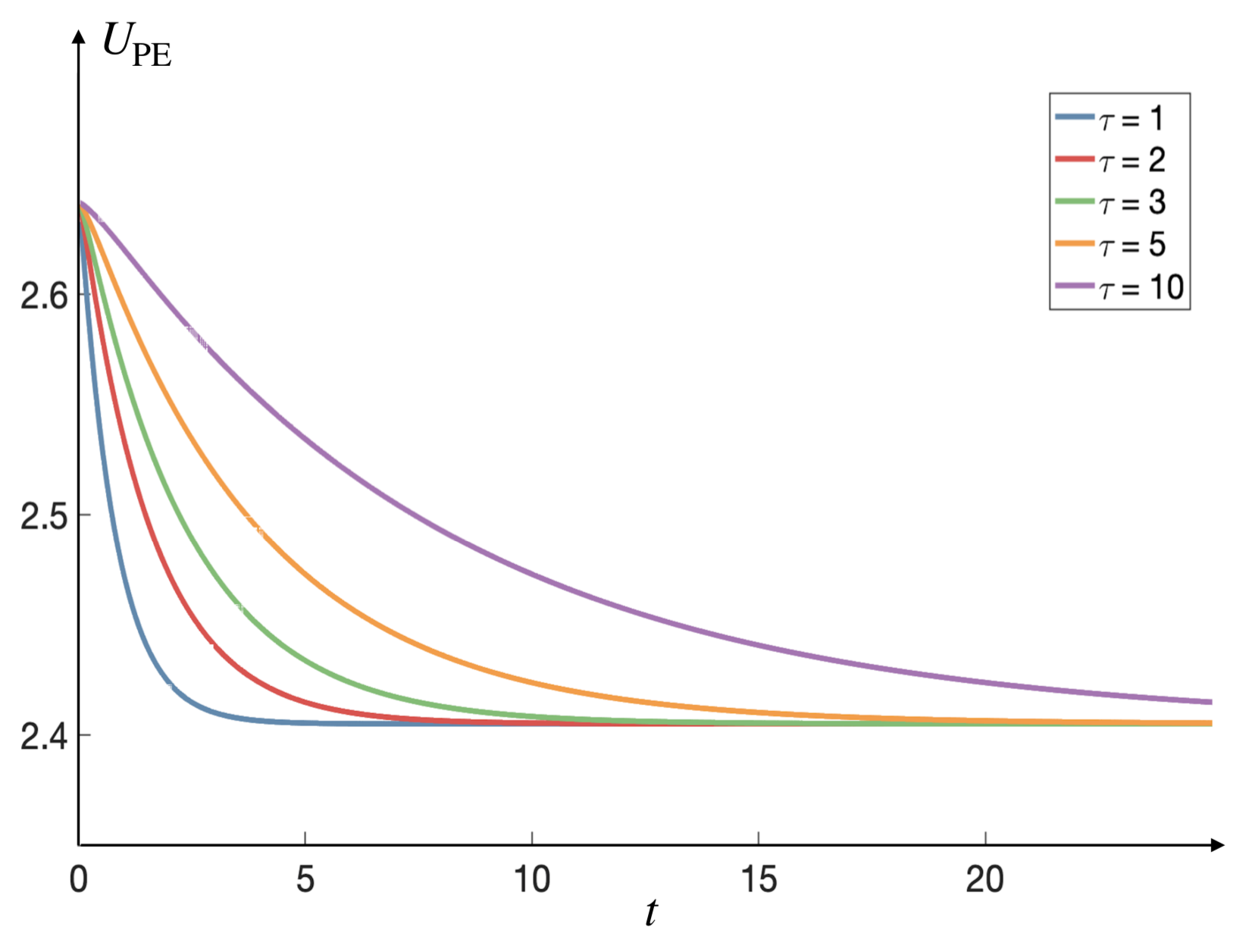}} \label{Fig6b} 
    \caption{(a) Velocity of the nucleus as a function of time for different values of permeability, ${\alpha}_{\circ}$, and a fixed value of elastic relaxation time, $\tau =1$. 
    At short ($t\to 0$) and long ($t\to \infty$) times the velocity asymptotes to $U_{\mathrm{v}}(\chi_1)$, and $U_\mathrm{B}({\alpha}_{\circ},\chi_1,\chi_2)$, respectively. 
    (b)  Velocity of the nucleus vs time for different values of elastic relaxation time $\tau$, and a fixed permeability ${\alpha}_{\circ} = 1$, when ${\chi}_1 = 1/3$, ${\chi}_2 = 2/3$, and ${\lambda}_1 = {\lambda}_{1}^{\prime} =1$.}
    \label{fig6}
\end{figure}
Figure \ref{fig6}(a) shows the velocity of the nucleus vs time for different values of ${\alpha}_{\circ}=\sqrt{1/\kappa}$ at $\tau=1$; and 
Fig. \ref{fig6}(b) shows the time-dependent nucleus velocity for different values of $\tau$, at ${\alpha}_{\circ} =1$. 
Also, we take ${\chi}_1 = 1/3$, ${\chi}_2 = 2/3$, ${\lambda}_1 = {\lambda}_{1}^{\prime} =1$.
As it can be seen, the initial velocity, $U_{\mathrm{PE}}(t=0)$, is independent  of $\alpha_{\circ}$ and $\tau$. 
At short times, the network deformation and hence its drag force against the fluid flow is small. 
Thus, the fluid permeates the network without resistance and the nucleus 
velocity asymptotes to the velocity obtained for a Newtonian fluid with $\chi=\chi_1$.
We confirm this analytically by using the relationship: 
$$\lim_{s\to \infty} s \,\tilde{U}_{\mathrm{PE}}(s)=\lim_{t\to 0}U_{\mathrm{PE}}(t)=U_{\mathrm{v}}(\chi_1).$$  

On the other hand, at long times network deformation reaches its stead-state and $\mathbf{v}_n \to U_{\mathrm{PE}}\hat{\mathbf{z}}$.
 In this limit, the momentum equation reduces to Brinkman equation $U_{\mathrm{PE}} \to U_{\mathrm{B}}(\chi_1,\chi_2,\alpha_{\circ})$. 
 We confirm this relationship by evaluating the following limit of $\tilde{U}_{\mathrm{PE}}$:
 $$ \lim_{s\to 0} s \, \tilde{U}_{\mathrm{PE}}(s)=\lim_{t\to \infty}U_{\mathrm{PE}}(t)=U_{\mathrm{B}}(\chi_1,\chi_2,\alpha_{\circ}).$$

At the intermediate times, the nucleus velocity approaches its steady-state value with an effective relaxation 
time that is a function of both $\tau$ and $\alpha_{\circ}$. An equation of form $y=y_i+(y_e-y_i)\exp(-t/\tau)$ 
fits the data presented in both figures very well. Given that the we know the limiting values of $U_{\mathrm{PE}}(t)$ at $t=0$ and $t \to \infty$, 
$U_{\mathrm{VE}}(t)$ can be approximated using the following general form:
\be
U_{\mathrm{PE}}(t)=U_{\mathrm{v}}(\chi_1)+\Big( U_{\mathrm{B}}(\chi_1,\chi_2,\alpha_\circ)-U_{\mathrm{v}}(\chi_1) \Big) \exp\left[-\frac{t}{\tau^\star_{\mathrm{PE}}(\tau,\alpha_{\circ},\chi_1,\chi_2)}\right],
\label{eq:taustar}
\ee 
where $\tau^\star_{\mathrm{PE}}$ is the relaxation timescale that depends on geometric parameters, as well as $\tau$ and $\alpha_{\circ}$. 

%\begin{figure}
  %\centerline{\includegraphics[width= \textwidth]{7}}% Images in 100% size
  %\caption{(a) Normalized poroelastic relaxation time as a function of permeability ${\alpha}_{\circ}$, for different values of elastic relaxation time $\tau$.  Inset: normalized poroelastic time scale for different size of cytoplasm, and different permeabilities.
  %(b)  The velocity of the nucleus vs. the velocity of the membrane, for periodic deformation of the membrane, and ${\lambda}_1 ={\lambda}_{1}^{\prime} =\cos(\omega t)$ in poroelastic cytoplasm. We assume ${\chi}_1 = 1/3$, ${\chi}_2 = 2/3$, and ${\lambda}_1 = {\lambda}_{1}^{\prime} =1$. 
%}
%\label{fig7}
%\end{figure}
\begin{figure}[!t]
    \centering
    \subfigure[]{\includegraphics[width=0.45\textwidth , height=6.7cm ]{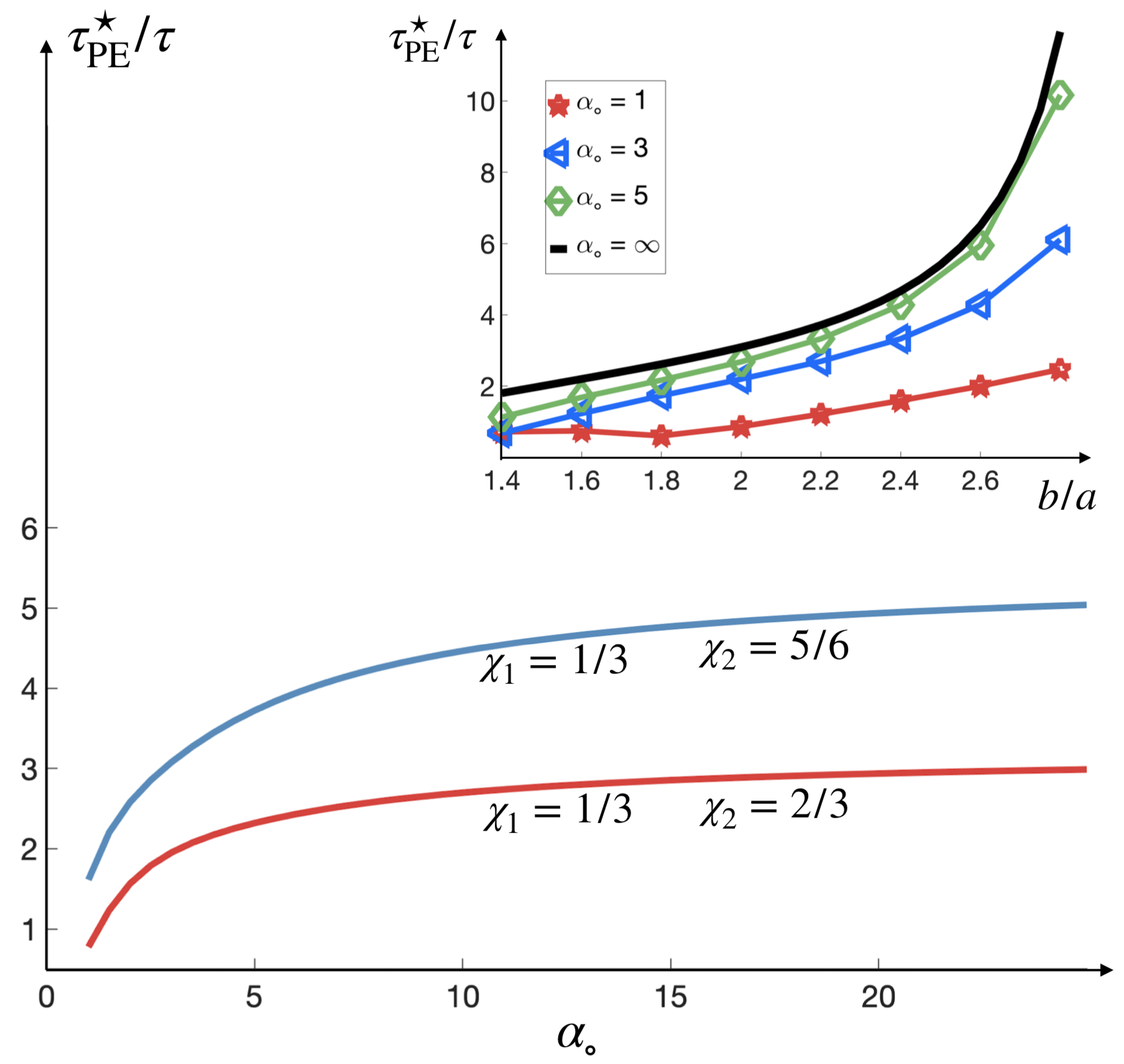}} \label{Fig7a}
    \subfigure[]{\includegraphics[width=0.45\textwidth , height=6.7cm]{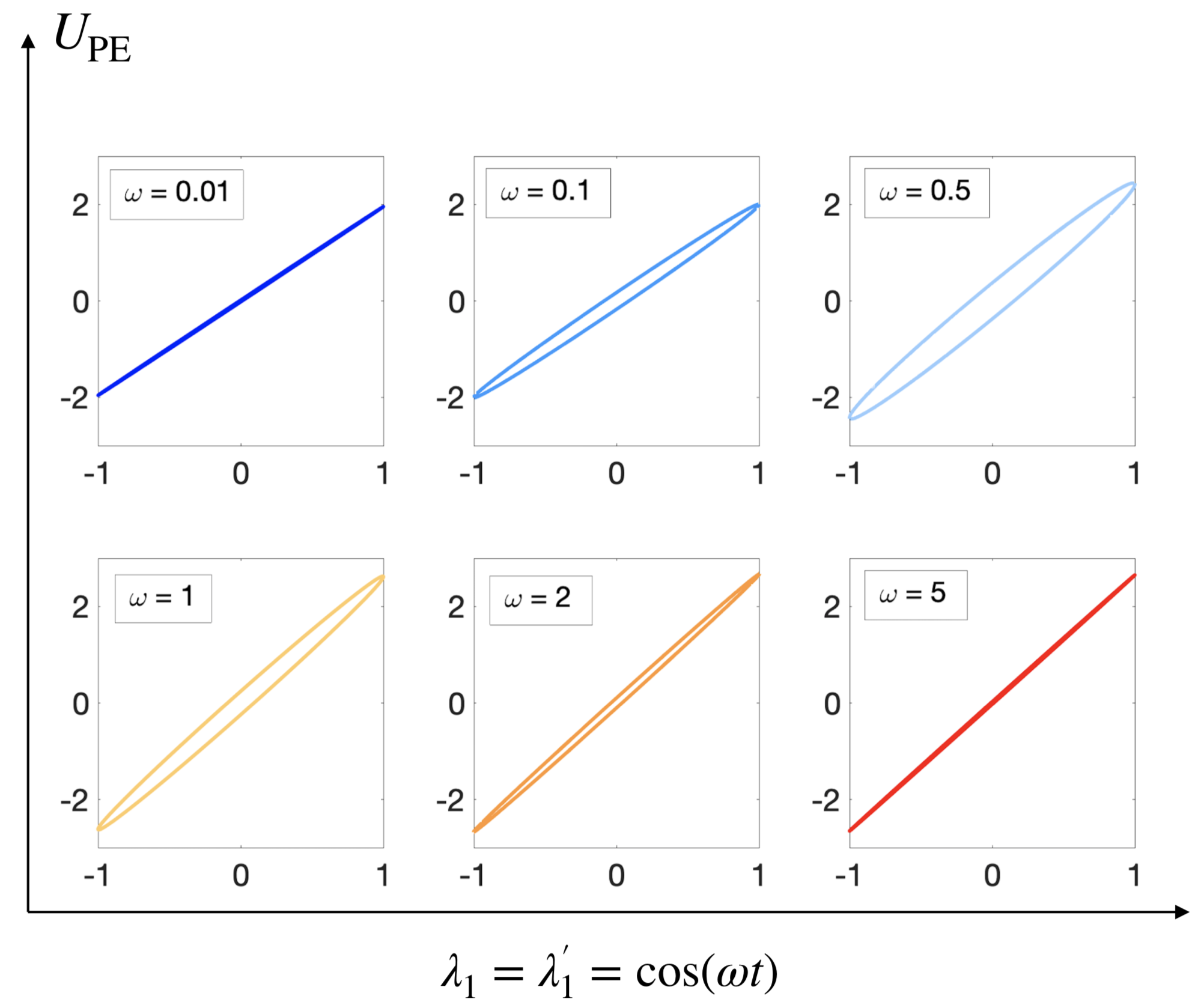}} \label{Fig7b}
    \subfigure[]{\includegraphics[width=0.45\textwidth , height=6.7cm]{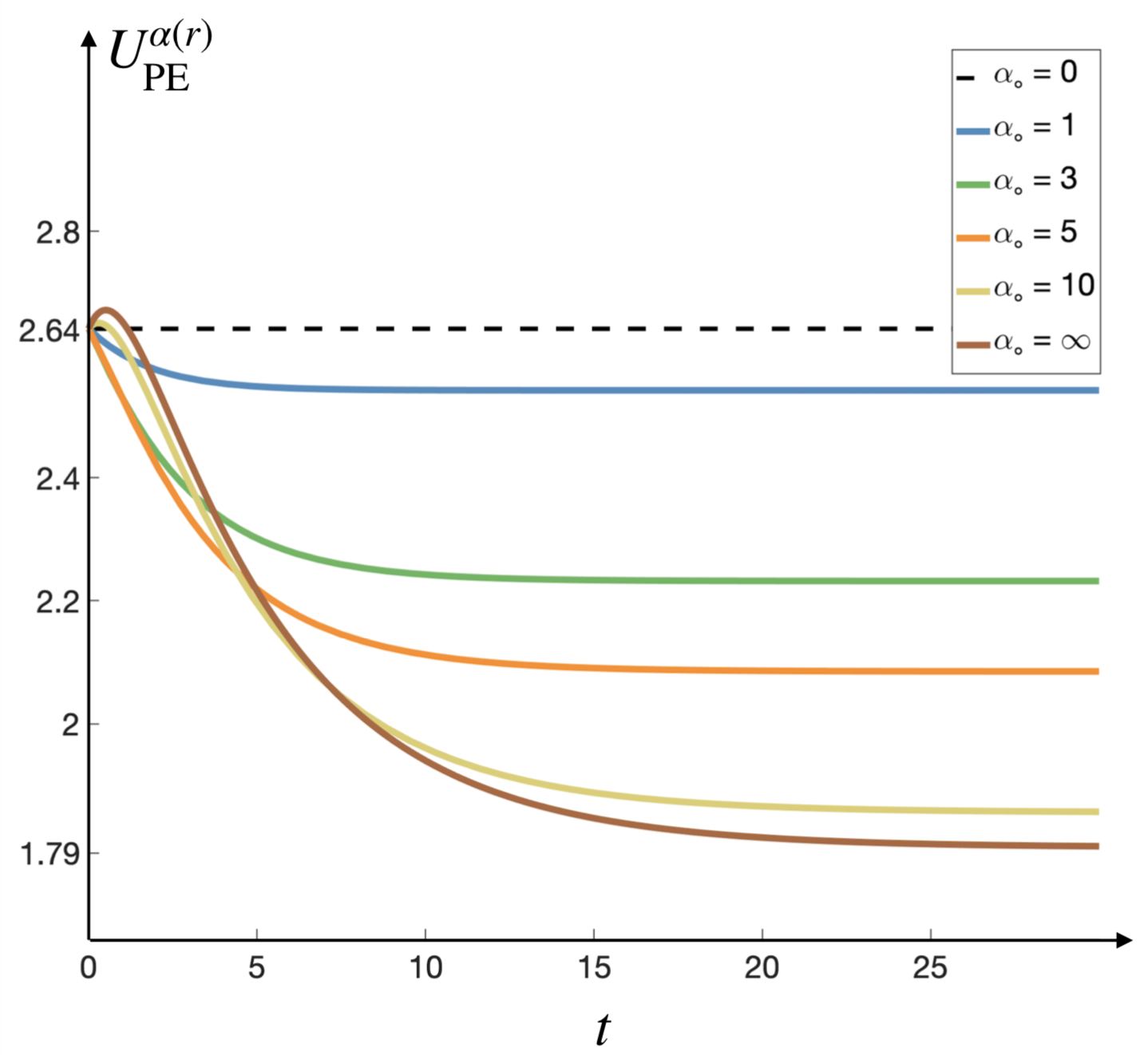}} \label{Fig7c}
    \subfigure[]{\includegraphics[width=0.45\textwidth , height=6.7cm]{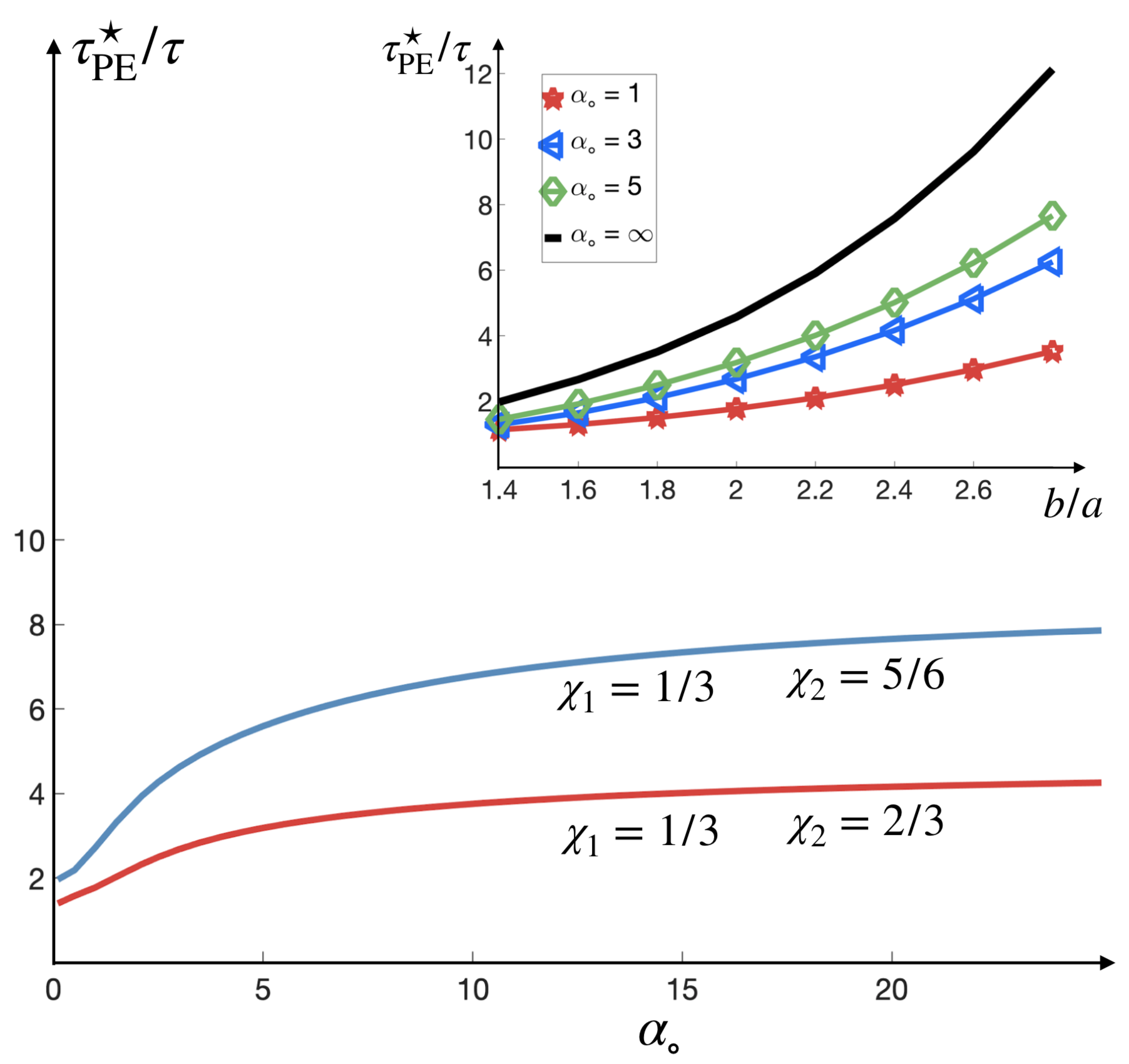}} \label{Fig7d}
    \caption{(a) Normalized poroelastic relaxation time, $\tau^\star_\text{PE}/\tau$, as a function of permeability ${\alpha}_{\circ}$, for two different values of poroelastic shell size, ${\chi}_2$, where $\tau^\star_\text{PE}$ is computed by fitting 
    Eq.\eqref{eq:taustar} on the results of $U_\text{PE}(t)$ vs $t$. Note that all the effective relaxation times, $\tau^\star$, for different values of $\tau$ collapse to a single curve, once they are normalized by $\tau$. Inset: normalized poroelastic time-scale as a function of the ratio of the shell to the nucleus radii, $b/a$, for different values of $\alpha_\circ$. The solid line correspond to the limit of $\alpha_\circ \to \infty$, given by Eq.\eqref{UpeInfPerm}.
    (b) Velocity of the nucleus vs. the membrane velocity, for a poroelastic shell when membrane deformations are periodic: ${\lambda}_1 ={\lambda}_{1}^{\prime} =\cos(\omega t)$. The results are reported for ${\chi}_1 = 1/3$, ${\chi}_2 = 2/3$, and different rate of deformation frequencies, $\omega$. 
     The curves at small and large frequencies, corresponding to long and short time-scales, asymptote to  linear curves signifying 
     nearly viscous behaviors. At intermediate frequencies 
    (time-scales), the nucleus velocity has a frequency-dependent phase lag ($\delta$), $U_{\mathrm{PE}}(t)=U_{\mathrm{PE}}^{0}\cos (\omega t+\delta)$, signifying a viscoelastic response. These phase lags result in tilted ellipse shapes, when plotting $U_{\mathrm{PE}}$ vs. $\lambda$ and $\lambda^\prime$.
     (c) Velocity of the nucleus as a function of time for a poroelastic cytoplasm with variable permeability, $\alpha = \frac{{\alpha}_{\circ}}{r}$, when ${\chi}_1 = 1/3$, ${\chi}_2 = 2/3$, and ${\lambda}_1 = {\lambda}_{1}^{\prime} =1$.
    (d) Normalized poroelastic relaxation time in variable permeability case, $\alpha (r) = \frac{{\alpha}_{\circ}}{r}$, as a function of ${\alpha}_{\circ}$, for two different shell outer radii, $\chi_2=2/3$, and $\chi_2=5/6$.  Inset: normalized poroelastic time-scale as a function of the ratio of the shell to the nucleus radii, $b/a$, for different permeabilities.}
    \label{fig7}
\end{figure}

Figure \ref{fig7}(a) shows the value of poroelastic timescale as a function of permeability. As it can be seen ${\tau}_{\mathrm{PE}}^{\star} / \tau$ 
monotonically increases (decreases) with $\alpha_{\circ}$ ($\kappa_{\circ}$) and 
approaches a constant that is independent of $\tau$ and only a function of $\chi_1$ and $\chi_2$ as the shell becomes impermeable ($\alpha_{\circ}\to \infty$). 
Since the exponential expression is an approximation to our solution, it is difficult to 
extract analytical expressions for ${\tau}_{\mathrm{PE}}^{\star}$ for general values of $\alpha_\circ$, $\tau$, $\chi_1$ and $\chi_2$. 
We can, however, obtain exact expressions for the nucleus velocity in the limit of nearly zero permeability, $\alpha_\circ\to\infty$. The expression 
is a combination of two exponential relaxations for constant values of $\lambda_1$ and ${\lambda}_{1}^{\prime}$:
\begin{align}
\begin{aligned}
U_{\mathrm{PE}} = U_{\mathrm{v}}({\chi}_2) + \left( {\lambda}_1 + {\lambda}_{1}^{\prime}  \right) \left[ f_1 ({\chi}_1 , {\chi}_2) {\mathrm{e}}^{-\frac{t}{\tau}} + f_2 ({\chi}_1 , {\chi}_2) {\mathrm{e}}^{-\frac{t}{\tau^{\star}} }  \right] , \qquad  \tau^{\star} =  \tau \left( \frac{15 {\chi}_{2}^{5} (1- {\chi}_{1}^{5} )}{(1- {\chi}_{2}^{5} )(2 {\chi}_{1}^{5} + 10 {\chi}_{1}^{2} {\chi}_{2}^{3} + 3 {\chi}_{2}^{5}    ) } \right), \label{UpeInfPerm}
\end{aligned}
\end{align} 
where $f_1 ({\chi}_1 , {\chi}_2)$ and $f_2 ({\chi}_1 , {\chi}_2)$ are lengthy geometric functions and are given in the Appendix~\ref{appB}.
One of these relaxation times is simply the relaxation time of the network, $\tau$, while the other one, $\tau^\star$, is proportional to $\tau$ and a function of geometry. 
Given that $b/a \ge 1$, $\tau^\star \ge \tau$. For example, for the choice of $\chi_1=1/3$ and $\chi_2=2/3$ we have $\tau^\star=3.1\tau$, which 
corresponds to the solid line shown in the inset of Fig.\ref{fig7}(a). 
As expected in the limit of $t \to \infty$, the nucleus velocity asymptotes to the velocity of a nucleus with radius $r=b$ in a Newtonian fluid, $U_{\mathrm{v}} ({\chi}_2)$.

Next, we compute the velocity of the nucleus for the case where the permeability is proportional to square of distance, $\kappa=\kappa_\circ r^2 \Rightarrow \alpha = \frac{{\alpha}_{\circ}}{r}$; see Figs.\ref{fig7}(c)-(d). 
Following the same steps as the case of constant permeability, 
we arrive at two equations for $\tilde{\mathbf{v}^+}$ and $\tilde{\mathbf{v}^-}$:
\begin{eqnarray}
&& \eta_{\circ} {\nabla}^2  {\tilde{\mathbf{v}}}^{+}  - \nabla {\tilde{p}}^+=\mathbf{0}, \\
&& \eta_{\circ} {\nabla}^2  {\tilde{\mathbf{v}}}^{-}  - \nabla {\tilde{p}}^- - \frac{\eta_{\circ}}{{\kappa}_{\circ} r^2} {\tilde{\mathbf{v}}}^{-}  =\mathbf{0},
\label{eq:PEvark}
\end{eqnarray} 
Similar to the case of constant permeability ${\tilde{\mathbf{v}}}^{+}$ satisfies Stokes equation, while the equation for ${\tilde{\mathbf{v}}}^{-}$ is the modified Brinkman equation with variable permeability, $\alpha=\alpha_{\circ}/r$. The analytical solution for the 
modified Brinkman equation is outlined in Appendix~\ref{appA}. The unknown coefficients in stream function solutions for Stokes and modified Brinkman equations are, then, determined by imposing BCs given by Eq.\eqref{eq:BCspm}. 

Figure \ref{fig7}(c) shows the variations of the nucleus velocity with time for different values of $\alpha_{\circ}$ when the permeability is spatially variant, $\alpha=\alpha_{\circ}/r$, 
and the choice of $\chi_1=1/3$, $\chi_2=2/3$ and $\tau=1$. 
The behavior is qualitatively similar to the case of constant permeability shown in 
Fig.\ref{fig6}(a). 
Variations of normalized effective relaxation time, $\tau^\star_\text{PE}/\tau$, 
vs $\alpha_{\circ}$ for a radially decaying $\alpha=\alpha_\circ/r$ is shown in Fig.\ref{fig7}(d) for 
two choices of $\chi_1$ and $\chi_2$. The inset figure in Fig.~\ref{fig7}c shows 
the variations of $\tau^\star_\text{PE}/\tau$ vs $\chi_2$ for different values of $\alpha_{\circ}$, when $\chi_1=1/3$. 
Again, in both cases, the results are qualitatively similar to 
the constant permeability case shown in Fig.\ref{fig7}(a). 
 
%\textcolor{blue}{Figure \ref{8} shows these normalized relaxation times ($\tau_1/\tau$ and $\tau_2/\tau$) 
%as a function of $\alpha_0$, when $\chi_1=1/3$ and $\chi_2=2/3$. }
%Note that the behavior is qualitatively similar to the case of constant permeability.
%\begin{figure}
 % \centerline{\includegraphics[width=  \textwidth]{8}}% Images in 100% size
 % \caption{ (a) Poroelastic shell with hinged boundary condition;  normal velocity and tangential velocity of the network equals the normal and tangential surface velocity modes. We assume ${\chi}_1 = 1/3$, ${\chi}_2 = 2/3$, and ${\lambda}_1 = {\lambda}_{1}^{\prime} =1$. (b) Poroelastic shell with normal velocity of the network equals the normal surface velocity modes, and tangential component of the network stress equal to zero. (c) Velocity of the inner sphere as a function of time for variable permeability, $\alpha = \frac{{\alpha}_{\circ}}{r}$.  }
%\label{fig8}
%\end{figure}
\begin{figure}[!htb]
    \centering
    \subfigure[]{\includegraphics[width=0.28\textwidth , height=4.5cm ]{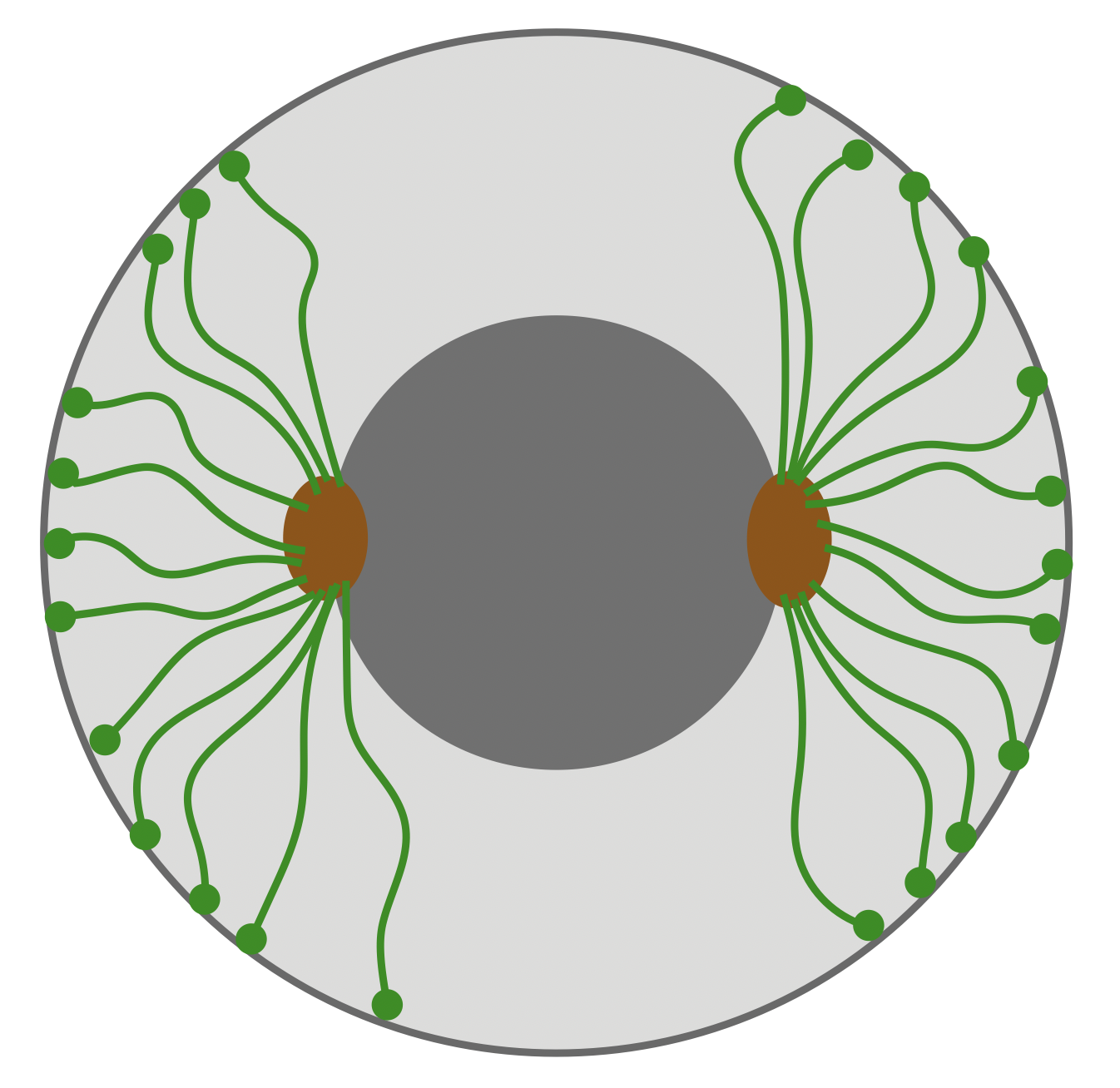}} \label{Fig8a}
    \qquad \qquad \qquad \qquad \qquad
    \subfigure[]{\includegraphics[width=0.28\textwidth , height=4.5cm]{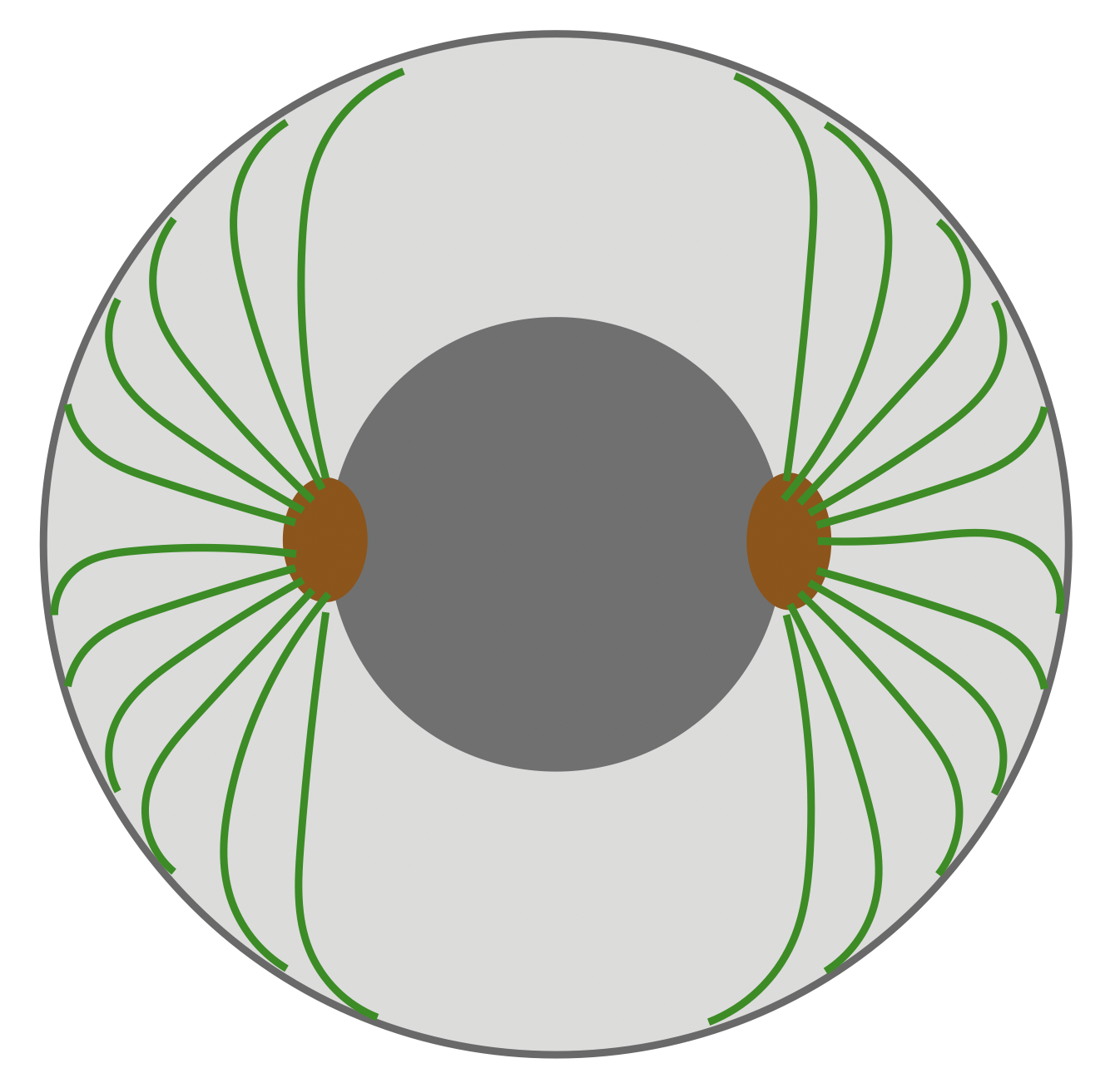}} \label{Fig8b}
    \subfigure[]{\includegraphics[width=0.45\textwidth , height=6.8cm]{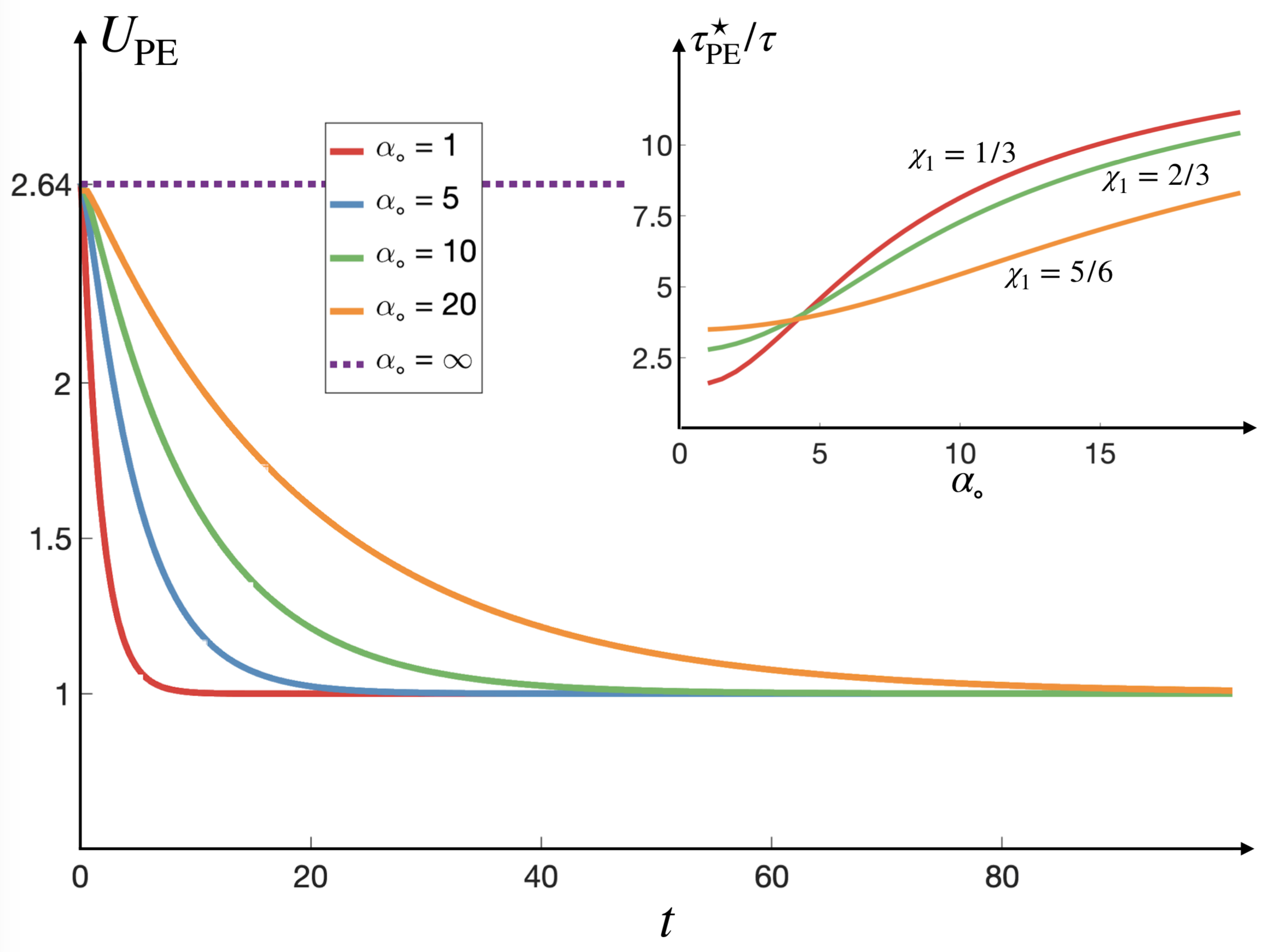}} \label{Fig8c}
    \hspace{0.2in}
    \subfigure[]{\includegraphics[width=0.45\textwidth , height=6.8cm]{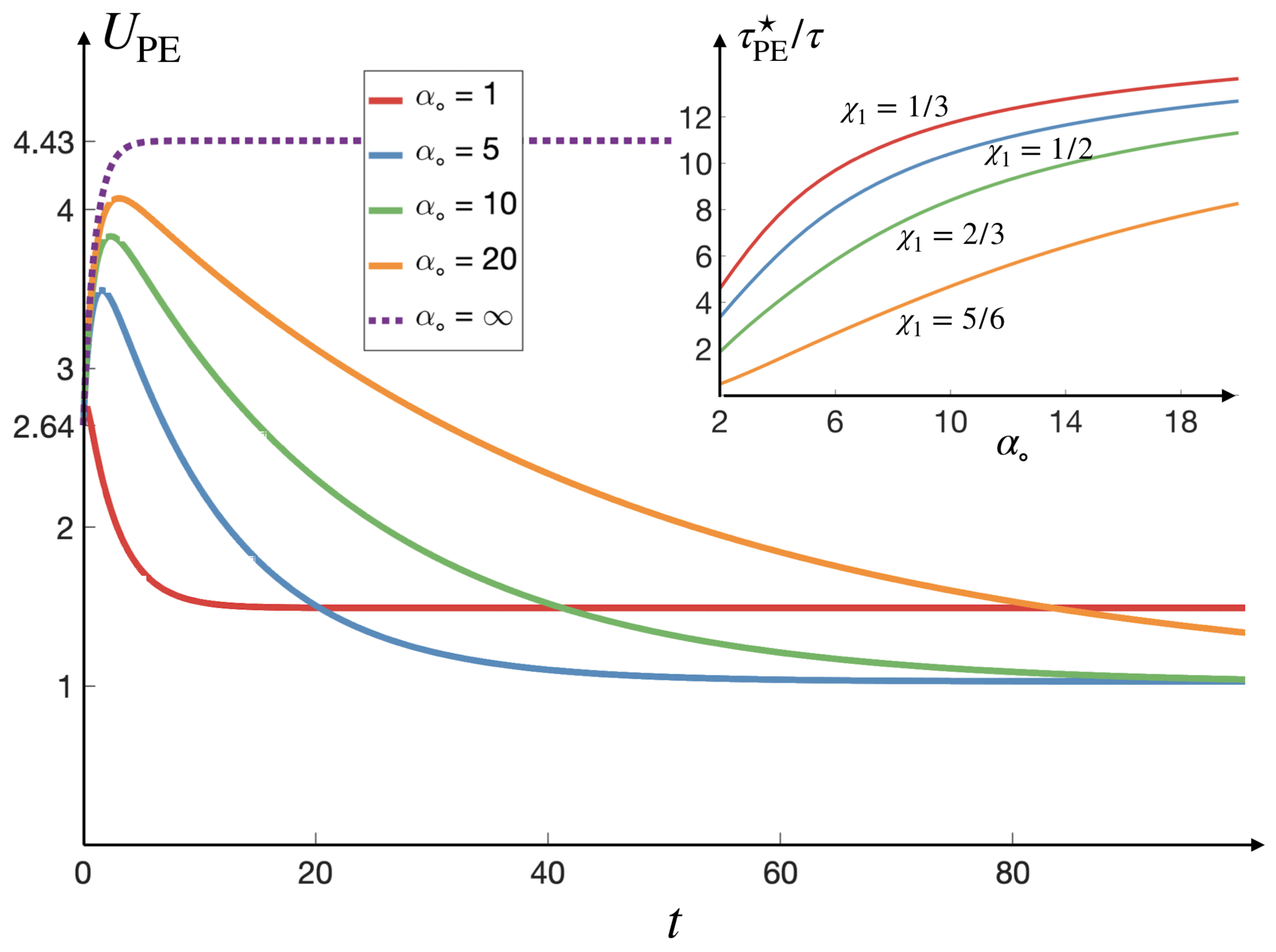}} \label{Fig8d}
    \caption{ Schematic representations of (a) scenario I, where the filament ends are mechanically trapped on the cell's cortical network. As a result, normal and tangential network displacements/velocities are equal to those of the cell cortex; and
    (b)  Scenario II, where the network filaments cannot penetrate the cell cortex (excluded volume), but can  bend and slide tangential to the boundary without resistance. Thus,  network displacement/velocity equals that of the cell cortex in the direction normal to the surface (no penetration), and the tangential component of the network stress vanishes (free sliding). 
    (c)  Velocity of the nucleus as a function of time in Scenario II and different values of $\alpha_{\circ}$, when ${\chi}_1 =1/3$, ${\lambda}_1 =  {\lambda}_{1}^{\prime} =1$, and $\tau =1$.   Inset: normalized effective relaxation time in Scenario II vs $\alpha_{\circ}$ for different nucleus radii, ${\chi}_1$.
    (d) Velocity of the nucleus as a function of time for a poroelastic spherical shell, where the BCs are the continuity of normal and tangential stress at the interface $r=b$, when $b\rightarrow c$. Inset: normalized effective poroelastic relaxation time as a function of permeability for different nucleus radii, ${\chi}_1$.  }
    \label{fig8}
\end{figure}
\subsubsection{Special case of \texorpdfstring{$\chi_2=1$}. } 
Thus far we have assumed that the poroelastic network that surrounds the nucleus does not fill the entire cytoplasm, $b<c$, \textit{i.e.}
the filaments that  make the network do not make mechanical contact with the cell cortex. 
This assumption may not hold in many physiological conditions. For example, during cell division many astral microtubules polymerize/depolymerize against cell cortex, 
and generate forces that are key for positioning the mitotic spindle \cite{nazockdast2017cytoplasmic}. 
In this limit, the fibers (microtubules) make contact with the cortex and any displacement of the cortex results in their deformations. 
Therefore, the force acting on the outer surface of the poroelastic network phase is no longer zero, 
and simply evaluating the expressions of the previous section in the limit of $b\to c$ will give the wrong results. 

What is the correct BCs for the outer surface of the network at $r=c$? 
The choice depends on the details of micromechanical interactions of the fibers with the outer boundary. 
We consider two scenarios. In scenario I, we assume 
the fibers remain mechanically attached to the cell cortex as it deforms. This scenario is shown schematically in Fig.\ref{fig8}(a). 
This can occur, for example, if the fiber ends are mechanically trapped in the dense cortical actin network. 
In this scenario, the network displacements will be equal to the cortical displacements.

In scenario II, we assume that the fibers cannot penetrate out of the cell boundary, \textit{i.e.} the network and the cortical displacements are equal in the direction normal to the cell boundary. However, we assume the fibers can bend and slide tangentially without any frictional resistance from 
the cellular boundary; see Fig.\ref{fig8}(b) for a schematic. The BCs for scenarios I and II are 
\begin{subequations}
\begin{align}
& \text{Scenario I}:  & &v_{n,r}=\sum_{n=1}^\infty \lambda_n P_n(\cos \theta), & &v_{n,\theta}=\sum_{n=1}^\infty \lambda^\prime_n V_n(\cos \theta) .\\
& \text{Scenario II}:  & &v_{n,r}=\sum_{n=1}^\infty \lambda_n P_n(\cos \theta), & &\sigma_{n,r\theta}=0 .
\end{align} 
\label{eq:BCPEspecial}
\end{subequations}
We, then, solve Eqs.\eqref{eq:PEpm} subject to no-slip BCs for the fluid and network phase at $r=a$ and BCs given in Eq.\eqref{eq:BCPEspecial} rewritten 
in terms $+$ and $-$ variables for both scenarios. We find that in scenario I, the nucleus velocity is simply given by $U=U_{\mathrm{v}} ({\chi}_1)$. 

The behavior is more complex and time-dependent in scenario II. 
Figure \ref{fig8}(c) shows the nucleus velocity vs time for different values of 
constant permeability $\alpha_{\circ}$ in scenario II, when $\lambda_1=\lambda^\prime_1=1$, $\chi_1=1/3$, and $\tau=1$. The inset shows the effective normalized relaxation time 
vs $\alpha_{\circ}$ for different nucleus radii ($\chi_1$). As a comparison, 
in Fig.\ref{fig8}(b) we show the results of Fig.\ref{fig8}(a), when we use the expressions for poroelastic shell in the limit of $b\to c$ to compute the nucleus velocity vs time. 
We can see that the predictions are qualitatively different from scenario II, underscoring the important role of micromechanical interactions and BCs in the limit where the network 
is in direct contact with the cell cortex. 
In reality the cortex/network interactions may likely be a combination of scenario I and II and the case of no contact (shell model), which also makes the nucleus velocity a combination of all these three ideal conditions.
\section{Summary}
Cell mechanics is a powerful tool for studying structural changes in different cellular processes. 
Cell populations and their mechanical properties are highly heterogeneous from cell to cell and from one location of the cell to the next. 
Thus, measurements need to be conducted for large number of cells. 
This constraint severely limits application of  several techniques such as AFM, and particle-tracking microrheology, specially for clinical applications. 
Using hydrodynamic forces in microfluidic platforms to deform cells is by far the highest throughput technique (thousands of cells per second), and 
most suitable for applications involving detecting and characterizing rare cell populations in clinical samples. 
Despite their remarkable success, the previous mechanical characterizations of cells in microchannels are limited in one fundamental way:
They largely use a single static measure of deformability as the metric for characterizing the mechanics. Therefore, it is very difficult to measure 
the contribution of different structural components of cell --including the cell nucleus, the cytoplasm and the cell membrane-- to the mechanics of the whole cell. 
This is particularly important as different processes and diseases are known target different cellular structures. 
Moreover, numerous studies of cells show that the rheology is time-dependent, which is not included in previous modeling studies. 

One major barrier in overcoming these limitations has been the lack of sufficient imaging resolution to probe the time-dependent deformations.  
In the past decade, there have been remarkable advancements in high speed fluorescent and optical imaging and their integration to microfluidic platforms. 
Due to these advancements we can now tag and track cellular boundaries
and different components within the cells as they flow through microchannels 
\cite{diebold2013digitally, jalai2016parallel, mikami2018ultrafast, rosendahl2018real} with high spatial and temporal resolution.  
Yet, there is no theoretical framework that enables utilizing this data to measure the time-dependent rheology of different cellular components. 

 This work introduces a computational method that makes use of this type of data to measure the rheology of 
 the cell cytoplasm and its immersed cytoskeleton over a wide range of time-scales.   
 As a model system, we considered a spherical cell with a centered spherical nucleus. 
 We assumed cell deformation and the nucleus displacements 
 are  sufficiently small that the nucleus remains centered and the cell remains spherical during the deformations.  
In these conditions the interior flow remains axisymmetric and the nucleus velocity can be computed analytically.  
 Bellow we summarize the main contributions of this method. 
\begin{enumerate}
\item \emph{Decoupling the rheology of the cell cortex from the cytoplasm:} 
As we discussed earlier, one standing challenge in studying cell mechanics is to determine the relative contribution of different cellular structure 
to the whole cell response measured in microfluidic experiments. 
We have proposed a novel method to measure the rheology of cytoplasm, without the need to include the mechanics of the cell cortex. 
The underlying assumption in all the calculations presented here is that velocity of the cell cortex is known from experiments through tracking the boundary. 
In this case, the internal flows and deformations can be fully determined by the cell cortex velocity, and the rheology of the cytoplasm, without the knowledge of the forces acting on the boundary and the mechanics of the cortex. 
\item \emph{Explicitly accounting for the cell nucleus, and using it as a microrheological probe:} 
Unlike previous studies where the cell interior was modeled as a single-phase material, 
we have modeled the cell interior as composition of the nucleus, and the cytosol with its immersed cytoskeleton, which we refer to as the cytoplasm. 
Considering that the nucleus is by far the stiffest structure inside the cell, we have modeled it as a rigid sphere. 

The nucleus moves in response to the cell's interior flows induced by cortical deformations. The relationship between nucleus and surface velocities is 
determined by the rheology of the cytoplasm. Here we assume that the nucleus displacement can be experimentally measured, which can then be used to 
characterize the rheology of the cytoplasm. 
Thus, including the nucleus not only provides a more accurate description of the cells interior, 
it can also be used as a microrheological probe to characterize the rheology of the cytoplasm. 

Nuclear mechanics can, however, be more complex, with these 
complexities playing a key role in many cellular processes. 
For example, growing evidence over the past decade suggest that the nucleus is a mechanosensory unit of the cell \cite{isermann2013nuclear}. 
Also, the nuclear composition and, thus, mechanics is known to be grossly affected by disease processes \cite{isermann2013nuclear,schreiber2013lamins}. 
Moreover, cell migration through tight extracellular environments may induce large deformations of the nucleus, which can have large effects on their migration time-scales \cite{wolf2013physical}. 
As we learn more about the multi-component nuclear organization, the composite 
mechanical models, like the ones presented here, can be constructed and applied to study the force transmission from the cell membrane, to the cytoplasm, nuclear envelope and nucleoplasm, and the effects of these forces on chromosome organization.  
 \item \emph{Accounting for the cell's structural heterogeneity:}
The density of cytoskeletal filaments such as microtubules  can vary greatly within the cell. To account for this, we have 
divided the cytoplasm into two volumes: a shell surrounding the nucleus, where filaments like microtubules are localized; 
and the remaining volume surrounding the shell, which we modeled as a Newtonian fluid. 
The shape and dimensions of these localized regions of filaments can, in practice, be visualized with the proper fluorescent tags 
for the relevant cytoskeletal filaments. We showed that including 
these structural heterogeneities can lead to qualitative changes in the nucleus dynamics.  

\item \emph{Assuming time-dependent rheology for the cytoplasm:}  
Previous modeling studies of cell deformation in microchannels model the cell as an elastic capsule or an elastic membrane filled with Newtonian fluid
\cite{mietke2015extracting, mokbel2017numerical}. These models do not capture the complex time-dependent response of the cell \cite{hoffman2009cell}. 
In this study,  we used four different rheological models for the shell surrounding the nucleus. 
In the first step, we modeled the entire cytoplasm as a Newtonian fluid. 
The nucleus velocity can be described in terms of the first modes of the cell's
radial and angular surface velocities, once they are decomposed into Legendre polynomials. 
The nucleus velocity in this case was only a function of the ratio of the nucleus to cell radii and independent of the viscosity of the cytoplasm; see Eq.\eqref{U1solid}.

Next, we modeled the shell as a general linear viscoelastic fluid. Taking Laplace transform of the time coordinate modifies momentum equation 
to Stokes equation with an $s-$dependent viscosity; see Eq.\eqref{eq:StokesLaplace}. The analytical solution to this equation provides an expression 
that relates the experimentally measured displacements of the nucleus and the cell cortex to time-dependent shear modulus; see Eq.\eqref{eq:VEmain}

As an example, we explored the predictions of the model when the fluid is described by Maxwell's equation. We considered constant and sinusoidal surface velocities, as 
analogs to creep/stress relaxation and linear oscillatory experiments in shear rheology. Note, however, that the formulation is applicable to any time-dependent form of surface velocities; see Fig.\ref{fig4}. We provided an analytical expression for the relaxation time of the nucleus velocity (see Eq.\eqref{eq:tauVE}),
as a function of the ratio of the radii of the nucleus and viscoelastic shell surrounding it to the cell radius, 
$\chi_1$ and $\chi_2$, and ratio of the viscosity of the filament phase to the cytosol, 
$\eta^\star$. 

Viscous and viscoelastic constitutive equations are most suitable for modeling force-free suspensions of cytoskeletal filaments. 
Since in most cytoskeletal assemblies the filaments are crosslinked, their motions are constrained. In such conditions, 
the cytoplasm is better described as a porous (rigid filaments) or a poroelastic (flexible filaments) medium.
We first modeled the shell as a porous medium and used Brinkman equation to model the fluid flows within it. 
These equation were solved analytically subject to appropriate BCs to give the nucleus velocity as a function of cytoplasm permeability, $\kappa$, 
and geometric parameters, $\chi_1$ and $\chi_2$; see Fig.\ref{fig5}. Thus, measuring the cell's surface and the nucleus 
velocities allows for computing the shell permeability.  

Next, we modeled the shell as a poroelastic medium composed of a viscoelastic network and a Newtonian fluid phases; see Eqs.\eqref{eq:PE}-\eqref{eq:PEC2}.
Through Laplace transform and a few simple change of variables, these equations were rewritten as Stokes and Brinkman equations in $s-$space;
see Eq.\eqref{eq:PEpm}. These equations were solved analytically to express the nucleus velocity as a function of the cell surface velocities, the geometric parameters $\chi_1$, $\chi_2$, and viscoelastic properties of the network, and its permeability $\kappa$. Similar to the case of viscoelastic shells, 
this expression can be used to compute the viscoelastic response of the network as well its permeability. 

As an example, we considered the case of a linear elastic network and 
studied the effect of geometrical factors $\chi_1$ and $\chi_2$, as well as the network relaxation time $\tau$ and permeability $\kappa$ on 
the predicted nucleus velocity; see Figs.\ref{fig7}-\ref{fig8}. Our results show that the relaxation timescale of the nucleus velocity is proportional to 
$\tau$ and changes significantly with $\alpha$, $\chi_1$ and $\chi_2$. 
\end{enumerate}

As discussed in the main text, relating the surface velocities to the nucleus velocity only allows us to compute the relaxation times of the cytoplasm, 
and not its viscosity. Determining the viscosity requires knowing the traction on the interior side of cell cortex, $\mathbf{f}_\text{I}$. 
On the other hand, since the viscosity of fluid in microchannels is known, we can compute the traction applied on the outer surface of the cell, 
$\mathbf{f}_\text{O}$, by solving Stokes equation (or Navier-Stokes in case inertial forces are important), subject to no-slip BCs on the surface. 
Force balance across the interface gives $\mathbf{f}_\text{I}+\mathbf{f}_\text{O}+\mathbf{f}_\text{M}=\mathbf{0}$, which means the 
the interior traction can only determined if the membrane traction is known. 
Thus, determining the viscosity of cytoplasm requires coupling of the mechanics of the membrane (cortex) and the interior flows.
We are currently working on extending our method to the case of the spherical cell moving in cylindrical channel, 
for which the exterior flow can be computed analytically.  

As mentioned, the aim of this study was to demonstrate the utility of such a modeling approach for characterizing 
the time-dependent rheology of the cytoplasm. Thus, we made several simplifying assumptions --which may not necessarily hold in the experimental and 
physiological conditions-- to obtain analytical expressions for the nucleus velocity. 
These assumptions include modeling the cell and the nucleus as spheres and that the cell deformations are sufficiently small that (i) the cell remains spherical, (ii) the nucleus remains roughly centered, and (iii) the material time derivatives appearing in viscoelastic constitutive equations can be 
approximated by partial derivatives: $D/Dt \approx \partial /\partial t$. 
These assumptions can certainly be relaxed, which would necessitate solving the equations numerically. 
We are currently pursuing this direction. 
\appendix
\section{Stream function solution for Stokes and Brinkman equations}\label{appA}
\subsection{Stokes equation stream function}\label{appA1}
 The Stokes and continuity equations are given by 
\begin{eqnarray}
&& \eta {\nabla}^2 \mathbf{v} - \nabla p =0, \qquad\qquad \mathrm{and} \qquad\qquad \nabla \cdot \mathbf{v} =0.  
\end{eqnarray}
Since the flow is axisymmetric, we have $v_\phi=0$ and $\partial/\partial \phi=0$ where $\phi \in [0,2\pi]$ and $\mathbf{v}(r,\theta)=(v_r,v_\theta)$ with $\theta \in [0,\pi]$. 
The velocity components are expressed in terms of Stokes stream function in spherical coordinate by
\be 
\begin{aligned}
{v}_{r} &=& - \frac{1}{r^2 \sin \theta} \frac{\partial {\psi}}{\partial\theta}   \\
 {v}_{\theta} &=& \frac{1}{r \sin\theta} \frac{\partial {\psi}}{\partial r}, 
\end{aligned}
\ee
where $\psi$ satisfies the following differential equation:
\be
\begin{aligned}
&& E^4 \psi = E^2 (E^2 \psi) = 0, \qquad\qquad E^2 = \frac{{\partial}^2}{\partial r^2} + \frac{\sin\theta}{r^2} \frac{\partial}{\partial \theta} \left( \frac{1}{\sin\theta} \frac{\partial}{\partial \theta} \right). 
\end{aligned}
\label{eq:psivel}
\ee
The general solution to the stream function is \cite{happel2012low}:
\be
\begin{aligned}
{\psi} (r,\theta) = \sum_{n=2}^{\infty} \left( A_n r^n + B_n r^{-n+1} + C_n r^{n+2} + D_n r^{-n+3} \right) {\mathfrak{G}}_{n}^{-1/2} (\cos \theta),  
\end{aligned}
\label{eq:psiStokes}
\ee
where ${\mathfrak{G}}_{n}^{-1/2} (\cos \theta)$ are Gegenbauer polinomials and are defined as $ {\mathfrak{G}}_{n}^{-1/2} (\cos \theta) := \frac{1}{2n-1} \left(  P_{n-2} (\cos\theta) - P_n (\cos\theta) \right)$, 
and $P_n(\cos \theta)$ is the Legendre polynomial of degree $n$.
Using Eq.\eqref{eq:psiStokes}, and ${\frac{\mathrm{d} {\mathfrak{G}}_{n}^{-1/2}(\cos\theta)}{\mathrm{d}(\cos\theta)}  }= - P_{n-1} (\cos\theta)$, modifies Eq.\eqref{eq:psivel} to
\begin{eqnarray}
 v_{r} &=& - \sum_{n=2}^{\infty} \left( A_n r^{n-2} + B_n r^{-n-1} + C_n r^n + D_n r^{-n+1}  \right) P_{n-1} (\cos\theta),  \\
 v_{\theta} &=& \sum_{n=2}^{\infty} \left(  n A_n r^{n-2} + (-n+1) B_n r^{-n-1} + (n+2) C_n r^n + (-n+3) D_n r^{-n+1} \right) \frac{{\mathfrak{G}}_{n}^{-1/2} (\cos\theta)}{\sin\theta}. 
\end{eqnarray}
Accordingly, the fluid stress components and the pressure are given by
\begin{eqnarray}
{\sigma}_{rr}^{\mathrm{H}} &=& 2 \eta \, \sum_{n=2}^{\infty} \Bigg[  (2-n) \frac{A_n}{r^{3-n}}  + (n+1) \frac{B_n}{r^{n+2}}  + \frac{-n^2+3n+1}{n-1} \frac{C_n}{r^{-n+1}} + \frac{n^2+n-3}{n} \frac{D_n}{r^{n}}    \Bigg]  P_{n-1} (\cos\theta),  \\
 {\sigma}_{r\theta}^{\mathrm{H}} &=& 2 \eta \sum_{n=2}^{\infty} \left[ \frac{2-n}{n-1} \frac{A_n}{r^{3-n}}  - \frac{n+1}{n} \frac{B_n}{r^{n+2}}  - \frac{n+1}{n} \frac{C_n}{r^{-n+1}}   -\frac{n-2}{n-1} \frac{D_n}{r^{n}}   \right]  \frac{\mathrm{d} P_{n-1} (\cos\theta)}{\mathrm{d}\theta},  \\
 p^{\mathrm{H}} &=& -2 \eta  \, \sum_{n=2}^{\infty} \Bigg[   \frac{2n+1}{n-1} \frac{C_n}{r^{-n+1}} + \frac{2n-3}{n} \frac{D_n}{r^{n}}    \Bigg]  P_{n-1} (\cos\theta). 
\end{eqnarray}

\subsection{Stream function for Brinkman equation with constant  permeability}\label{appA2}
The Brinkman equation for a porous medium and the continuity equation are
\begin{eqnarray}
\eta {\nabla}^2 \mathbf{v}  - \eta {\alpha}^2 \mathbf{v} - \nabla p = 0,   \qquad\qquad \mathrm{and} \qquad\qquad \nabla \cdot \mathbf{v} =0  , 
\end{eqnarray} 
where $\alpha = \frac{1}{\sqrt{\kappa}}$, and $\kappa$ is the permeability. 
The stream function for constant permeability, $\alpha = {\alpha}_{0}$, satisfies the following differential equation:
\begin{eqnarray}
&& E^4  {\psi}^{\star} - {\alpha}_{\circ}^2 (E^2  {\psi}^{\star}) = 0.  
\end{eqnarray}
Solutions of the above equation may be obtained by setting 
${\psi}^{\star} = {\psi}^{(1)} + {\psi}^{(2)}$, where $\psi_1$ and 
$\psi_2$ satisfy the following differential equations
\begin{subequations}
\begin{align}
&(E^2 - {\alpha}_{\circ}^{2}) {\psi}^{(1)} =0\\
& (E^2 - {\alpha}_{\circ}^{2}) {\psi}^{(2)} =W \\
& E^2 W=0.
\end{align}
\end{subequations}
The solution for ${\psi}^{(2)}$ is ${\psi}^{(2)} = -\frac{1}{{\alpha}_{\circ}^2} W$, which is $\left( a_n r^{n} + b_n r^{-n+1} \right) {\mathfrak{G}}_n (\cos\theta)$. The solution for ${\psi}^{(1)}$ is obtained by separation of variables, ${\psi}^{(1)} = R(r) Z(\xi)$, which upon substitution yields $(\xi = \cos(\theta))$
\begin{subequations}
\begin{eqnarray}
 && r^2 \frac{{\mathrm{d}}^2 R}{\mathrm{d} {r}^2} - n(n+1) R - {\alpha}_{\circ}^2 r^2 R =0,    \\
 &&  (1- {\xi}^2) \frac{{\mathrm{d}}^2 Z}{\mathrm{d} {\xi}^2} + n(n-1) Z = 0.   
\end{eqnarray}
\end{subequations}
The latter equation is Gegenbauer equation with degree $-\frac{1}{2}$. By change of variable $R = \sqrt{r} R_1$, the differential equation for $R_1$ transforms to the modified Bessel differential equation:
\begin{eqnarray}
r^2 \frac{{\mathrm{d}}^2 R_1}{\mathrm{d} {r}^2} + r \frac{\mathrm{d} R_1}{\mathrm{d} r}  - \left( {(n-\frac{1}{2})}^2 + {\alpha}_{\circ}^2 r^2 \right) R_1  =0. 
\end{eqnarray}
Thus, the general solution for the stream function of Brinkman fluids is given by
\begin{eqnarray}
 {\psi}^{\star}(r,\theta) &=& \sum_{n=2}^{\infty} \left[ {\tilde{A}}_{n} r^n + {\tilde{B}}_n r^{(-n+1)} + {\tilde{C}}_n y_n ({\alpha}_{\circ} r) + {\tilde{D}}_n y_{-n} ({\alpha}_{\circ} r)  \right] {\mathfrak{G}}_{n}^{-1/2} (\cos \theta). 
\end{eqnarray}
Here, ${\alpha}_{\circ} = 1/\sqrt{\kappa}$, and $y_n ({\alpha}_{\circ} r)$ and $y_{-n} ({\alpha}_{\circ} r)$ are related to the modified Bessel function as
\begin{eqnarray}
y_n ({\alpha}_{\circ} r) = \sqrt{\frac{\pi {\alpha}_{\circ} r}{2}} {{\alpha}_{\circ}}^{n-\frac{1}{2}} \,\, {\mathrm{I}}_{n-\frac{1}{2}} ({\alpha}_{\circ} r) , \qquad\qquad  y_{-n} ({\alpha}_{\circ} r) = \sqrt{\frac{\pi {\alpha}_{\circ} r}{2}} {{\alpha}_{\circ}}^{n-\frac{1}{2}} \,\, {\mathrm{I}}_{-(n-\frac{1}{2})} ({\alpha}_{\circ} r).
\end{eqnarray} 
For $n=2$, $y_2 ({\alpha}_{\circ} r) = {\alpha}_{\circ} \cosh ({\alpha}_{\circ} r) - \frac{1}{r} \sinh ({\alpha}_{\circ} r), $ and $ y_{-2} ({\alpha}_{\circ} r) =  {\alpha}_{\circ} \sin ({\alpha}_{\circ} r) - \frac{1}{r} \cosh ({\alpha}_{\circ} r) $.
The pressure and the tangential stress are:
\begin{eqnarray}
&& p = \eta {{\alpha}_{\circ}}^2 \sum_{n=2}^{\infty} \left( \frac{1}{n-1} {\tilde{A}}_{n} r^{n-1} - \frac{1}{n} {\tilde{B}}_{n} r^{-n} \right) P_{n-1} (\cos\theta)   \\
&&  {\sigma}_{r \theta} = 2 \eta \sum_{n=2}^{\infty} \Bigg( n(n-2) {\tilde{A}}_{n} r^{n-3} + (n^2 -1) {\tilde{B}}_{n} r^{-n-2} +  \frac{{\tilde{C}}_{n}}{2 r^3} \left( n(n-1) y_n ({\alpha}_{\circ} r) -2r y_{n}^{\prime} ({\alpha}_{\circ} r) + r^2 y_{n}^{\prime\prime} ({\alpha}_{\circ} r)  \right)  +  \nn \\
&& \qquad\qquad\qquad\qquad\qquad +   \frac{{\tilde{D}}_{n}}{2 r^3} \left( n(n-1) y_{-n} ({\alpha}_{\circ} r) -2r y_{-n}^{\prime} ({\alpha}_{\circ} r) + r^2 y_{-n}^{\prime\prime} ({\alpha}_{\circ} r) \right) \Bigg) \frac{ {\mathfrak{G}}_{n}^{-1/2} (\cos \theta)}{\sin\theta}   
\end{eqnarray}

\subsection{Stream function for Brinkman equation with variable permeability }\label{appA3}
For variable permeability, $\alpha(r)$, the stream function  satisfies the fourth order differential equation:
\begin{eqnarray}
&& E^4  {\psi}^{\star} - {\alpha}^2 (r) \left( E^2 + \frac{2}{\alpha (r)} \frac{\mathrm{d}\alpha}{\mathrm{d}r} \frac{\mathrm{d}}{\mathrm{d}r}  \right)  {\psi}^{\star} = 0.  
\end{eqnarray}
Upon separation of variables, ${\psi}^{\star} = R(r)Z(\xi)$, one finds that the differential equation for $Z(\xi)$ and its solution are identical to the case of constant permeability, and the differential equation for $R(r)$ is
\be
   r^4  \frac{{\mathrm{d}}^4 R}{\mathrm{d} {r}^4} - \left( {\alpha}^2 r^2 + 2n(n-1)  \right) r^2  \frac{{\mathrm{d}}^2 R}{\mathrm{d} {r}^2}  - \left( 2\alpha r^3 \frac{\mathrm{d}\alpha}{\mathrm{d}r} -4n(n-1)  \right) r \frac{\mathrm{d} R}{\mathrm{d} r} + n(n-1) \left( {\alpha}^2 r^2 +n^2 -n -6  \right) R =0 .  
\ee
For the special case of $\alpha = \frac{{\alpha}_{\circ}}{r}$, the differential equation simplifies to:
\begin{eqnarray}
&&   r^4  \frac{{\mathrm{d}}^4 R}{\mathrm{d} {r}^4} - \left( {\alpha}_{\circ}^2  + 2n(n-1)  \right) r^2  \frac{{\mathrm{d}}^2 R}{\mathrm{d} {r}^2}  + \left( 2 {\alpha}_{\circ}^2 +4n(n-1)  \right) r \frac{\mathrm{d} R}{\mathrm{d} r} + n(n-1) \left( {\alpha}_{\circ}^2  +n^2 -n -6  \right) R =0 .  
\end{eqnarray}
Changing to variable $r={\mathrm{e}}^{\rho}$, we get:
\begin{align}
   R^{\prime\prime\prime\prime}(\rho) -& 6 R^{\prime\prime\prime}(\rho) + (-2n(n-1)+11 - {\alpha}_{\circ}^2) R^{\prime\prime}(\rho)+ \\
& (3 {\alpha}_{\circ}^2 + 6n(n-1) -6) R^{\prime} (\rho) + n(n-1) ({\alpha}_{\circ}^2 + n^2 -n-6) R (\rho) =0 .  \nn
\end{align}
The general solution of the above differential equation is of the form ${\mathrm{e}}^{m \rho} = r^m$, where $m$ satisfies the following quartic equation:
\begin{eqnarray}
&&   m^4 -6 m^3 + (-2n(n-1)+11 - {\alpha}_{\circ}^2) m^2 + (3 {\alpha}_{\circ}^2 + 6n(n-1) -6) m  + n(n-1) ({\alpha}_{\circ}^2 + n^2 -n-6)  =0 .  \nn
\end{eqnarray}
Hence, the stream function is
\begin{eqnarray}
&&   {\psi}^{\star} = \sum_{n=2}^{\infty} \left( A r^{m_1} + B r^{m_2} + C r^{m_3} + D r^{m_4}  \right) {\mathfrak{G}}_n (\cos \theta),  
\end{eqnarray}
where $m_i , (i=1,2,3,4)$ are roots of the quartic equation.  For ${\alpha}_{\circ} =0$, we recover Stokes solutions, $r^{n}$, $r^{-n+1}$, $r^{n+2}$, and $r^{-n+3}$. 

For $n=2$, the solutions of quartic equation are
\begin{eqnarray}
&& m_{1,2,3,4} = \frac{1}{2} \left( 3 \pm \sqrt{13 + 2 {\alpha}_{\circ}^2 \pm 2 \sqrt(36 - 4  {\alpha}_{\circ}^2+  {\alpha}_{\circ}^4 )}   \right).   
\end{eqnarray}
The velocity components are
\begin{subequations}
\begin{eqnarray}
&& v_r = -\cos \theta \left( A r^{m_1 -2} + B r^{m_2 -2} + C  r^{m_3 -2} + D  r^{m_4 -2}  \right)  ,   \\
&&  v_{\theta} = \frac{1}{2} \sin \theta \left( m_1 A  r^{m_1 -2} + m_2 B  r^{m_2 -2}  + m_3 C  r^{m_3 -2}  + m_4 D  r^{m_4 -2} \right), 
\end{eqnarray}
\end{subequations}
and the pressure and tangential stress components are
\begin{subequations}
\begin{eqnarray}
&& p = \eta \cos \theta \Bigg[ \left( \frac{m_1^2 - m_1 -2 - {\alpha}_{\circ}^2}{m_1 -3} \right) A \, r^{m_1 -3} +  \left( \frac{m_2^2 - m_2 -2 - {\alpha}_{\circ}^2}{m_2 -3} \right) B \, r^{m_2 -3} +  \left( \frac{m_3^2 - m_3 -2 - {\alpha}_{\circ}^2}{m_3 -3} \right) C \, r^{m_3 -3} +  \nn \\
&& \qquad\qquad\qquad\qquad\qquad\qquad \qquad\qquad\qquad\qquad\qquad\qquad\qquad\qquad  + \left( \frac{m_4^2 - m_4 -2 - {\alpha}_{\circ}^2}{m_4 -3} \right) D \, r^{m_4 -3}    \Bigg]   ,   \\
&&  {\sigma}_{r \theta} = \eta \sin \theta \Bigg[  \left( \frac{3 m_1^2 - 11 m_1 +10 - {\alpha}_{\circ}^2}{m_1 -3} \right) A \, r^{m_1 -3} +  \left( \frac{3 m_2^2 - 11 m_2 +10 - {\alpha}_{\circ}^2}{m_2 -3} \right) B \, r^{m_2 -3} +  \nn \\   
&& \qquad\qquad\qquad\qquad +  \left( \frac{3 m_3^2 - 11 m_3 +10 - {\alpha}_{\circ}^2}{m_3 -3} \right) C \, r^{m_3 -3} +  \left( \frac{3 m_4^2 - 11 m_4 +10 - {\alpha}_{\circ}^2}{m_4 -3} \right) D \, r^{m_4 -3}  \Bigg] . 
\end{eqnarray}
\end{subequations}

\section{Functions $f_1 ({\chi}_1 , {\chi}_2)$ and $f_2 ({\chi}_1 , {\chi}_2)$} \label{appB}
The functions $f_1 ({\chi}_1 , {\chi}_2)$ and $f_2 ({\chi}_1 , {\chi}_2)$ that appear in equation~\eqref{UpeInfPerm} are:
\begin{align}
 f_1 =  \frac{10}{3} \left( \frac{ {({\chi}_1 - {\chi}_2)}^3  (4 {\chi}_{1}^{2} {\chi}_{2}^{2} + 7 {\chi}_{1} {\chi}_{2}^{3} + 4 {\chi}_{2}^{4} )}{{\Delta}_1  }  \right), \qquad
 f_2 =   -\frac{5}{3} \left( \frac{ ({\chi}_1 - {\chi}_2) ( {\chi}_1^4 + {\chi}_{1}^{3} {\chi}_2 + {\chi}_{1}^{2} {\chi}_{2}^{2} + {\chi}_{1} {\chi}_{2}^{3} + {\chi}_{2}^{4}    ) {\Delta}_2  } { {\Delta}_3 }  \right) ,   
\end{align}
with

\begin{flalign} 
 {\Delta}_1 &= 13 {\chi}_{1}^{5} {\chi}_{2}^{5} + 13 {\chi}_{1}^{4} {\chi}_{2}^{6} + 13 {\chi}_{1}^{3} {\chi}_{2}^{7} + 2 {\chi}_{1}^{5} + 2 {\chi}_{1}^{4} {\chi}_{2} + 2 {\chi}_{1}^{3} {\chi}_{2}^{2} + 3 {\chi}_{1}^{2} {\chi}_{2}^{8} + 3 {\chi}_{1} {\chi}_{2}^{9} + 12 {\chi}_{1}^{2} {\chi}_{2}^{3} + 12 {\chi}_{1} {\chi}_{2}^{4}, && \nn \\
 {\Delta}_2 &= 8( {\chi}_{2}^{6} + {\chi}_{2}^{5} + {\chi}_{2}^{4} + {\chi}_{2}^{3} + {\chi}_{2}^{2}  ) -3 {\chi}_{1}^{4} {\chi}_{2}^{7} -3 {\chi}_{1}^{4} {\chi}_{2}^{6} + 10  {\chi}_{1}^{4} {\chi}_{2}^{5} + 10  {\chi}_{1}^{4} {\chi}_{2}^{4} + 10  {\chi}_{1}^{4} {\chi}_{2}^{3} -2  {\chi}_{1}^{4} {\chi}_{2}^{2} -2  {\chi}_{1}^{4} {\chi}_{2} -3  {\chi}_{1}^{3} {\chi}_{2}^{8}  &&\nn \\
& \qquad -3  {\chi}_{1}^{3} {\chi}_{2}^{7} + 10  {\chi}_{1}^{3} {\chi}_{2}^{6} + 10  {\chi}_{1}^{3} {\chi}_{2}^{5} + 10  {\chi}_{1}^{3} {\chi}_{2}^{4} -2  {\chi}_{1}^{3} {\chi}_{2}^{3} -2  {\chi}_{1}^{3} {\chi}_{2}^{2} -21  {\chi}_{1}^{5} {\chi}_{2}^{6} -21  {\chi}_{1}^{5} {\chi}_{2}^{5} -8  {\chi}_{1}^{5} {\chi}_{2}^{4} -8  {\chi}_{1}^{5} {\chi}_{2}^{3}  && \nn \\
 & \qquad -8  {\chi}_{1}^{5} {\chi}_{2}^{2} -2  {\chi}_{1}^{5} {\chi}_{2} -2  {\chi}_{1}^{5}  + 3  {\chi}_{1}^{2} {\chi}_{2}^{7} + 3  {\chi}_{1}^{2} {\chi}_{2}^{6} + 3  {\chi}_{1}^{2} {\chi}_{2}^{5} -10  {\chi}_{1}^{2} {\chi}_{2}^{4} -10  {\chi}_{1}^{2} {\chi}_{2}^{3} + 2  {\chi}_{1}^{2} {\chi}_{2}^{2} + 2  {\chi}_{1}^{2} {\chi}_{2} + 2  {\chi}_{1}^{2} + 3  {\chi}_{1} {\chi}_{2}^{8} && \nn \\
 & \qquad  + 3  {\chi}_{1} {\chi}_{2}^{7}  + 3  {\chi}_{1} {\chi}_{2}^{6} -10  {\chi}_{1} {\chi}_{2}^{5}  -10  {\chi}_{1} {\chi}_{2}^{4} + 2  {\chi}_{1} {\chi}_{2}^{3} + 2  {\chi}_{1} {\chi}_{2}^{2} + 2  {\chi}_{1} {\chi}_{2}  ,           &&       \nn \\
{\Delta}_3 &= (1- {\chi}_{1}^{5} ) ({\chi}_{2}^{4} + {\chi}_{2}^{3} + {\chi}_{2}^{2} + {\chi}_{2} +1 ) {\Delta}_1 . && \nn
\end{flalign}

\bibliographystyle{unsrt}
\bibliography{MybibE}

\end{document}